\documentclass[%
superscriptaddress,
nofootinbib,
twocolumn]{revtex4-1}

\usepackage{amssymb, amsmath, bm, dcolumn, epsf, graphicx, latexsym, slashed, simplewick}
\usepackage[utf8]{inputenc}

\usepackage{graphicx}
\usepackage{dcolumn}
\usepackage{bm}
\usepackage{xcolor}
\usepackage{bbold}
\usepackage{soul}
\usepackage[usestackEOL]{stackengine}
\usepackage{color}

\usepackage{amsmath}
\usepackage{amssymb}
\usepackage[normalem]{ulem}
\usepackage{comment}
\usepackage{pythonhighlight}  
\usepackage{hyperref}

\begin{document}

\preprint{APS/123-QED}

\title{The Physics of Machine Learning:\\An Intuitive Introduction for the Physical Scientist}
\author{Stephon Alexander}
\affiliation{Brown Theoretical Physics Center, Providence, RI 02912, USA}
\affiliation{Department of Physics, Brown University, Providence, RI 02912, USA}

\author{Sarah Bawabe}
\affiliation{Brown Theoretical Physics Center, Providence, RI 02912, USA}
\affiliation{Department of Physics, Brown University, Providence, RI 02912, USA}

\author{Batia Friedman-Shaw}
\affiliation{Brown Theoretical Physics Center, Providence, RI 02912, USA}
\affiliation{Department of Physics, Brown University, Providence, RI 02912, USA}

\author{Michael W. Toomey}
\affiliation{Brown Theoretical Physics Center, Providence, RI 02912, USA}
\affiliation{Department of Physics, Brown University, Providence, RI 02912, USA}

\begin{abstract}
    This article is intended for physical scientists who wish to gain deeper insights into machine learning algorithms which we present via the domain they know best, physics. We begin with a review of two energy-based machine learning algorithms, Hopfield networks and Boltzmann machines, and their connection to the Ising model. This serves as a foundation to understand the phenomenon of learning more generally. Equipped with this intuition we then delve into additional, more ``practical,'' machine learning architectures including feedforward neural networks, convolutional neural networks, and autoencoders.  We also provide code that explicitly demonstrates training a neural network with gradient descent.
\end{abstract}

\maketitle

\section{Introduction}

With the advent of new state-of-the-art algorithms and faster hardware, the application of machine learning (ML) in the sciences has seen rapid growth. Now more than ever, the need for clear and concise introductions to the topic are critical. The main point of attraction for ML algorithms stem from the power these architectures have when applied to high dimensional problems -- an established example is the classification of images of cats and dogs. Given their unique structure, ML architectures are often described as a \textit{black box} -- data goes in, some magical process or computation occurs in the architecture and, more often than not, a correct results is returned. We contend that the nature and dynamics is not as mysterious as it seems at first, something which we hope becomes clear through the lens of theoretical physics.

While there are many primers on machine learning in the literature, they are either strictly related to ML or intended for other scientific fields such as epidemiology~\cite{ML_epidemiology}, ecology~\cite{ANNs_ecology}, and cognitive or neuroscience~\cite{anderson1995introduction}. There are, however, few intended for physical scientists. Thus, given our audience, we will leverage the fact that the foundations of many ML algorithms are based in statistical mechanics. Indeed, many important architectures in machine learning are inspired by the Ising model~\cite{hertz_1991}. In this way, we will show how the Ising model can be viewed as a ``standard model'' for many ML architectures. We will introduce machine learning economically, that is, by introducing only topics directly related to the Ising model as our starting point, and then branching out from those to introduce a few generally common neural network topics, such as feedforward neural networks, convolutional neural networks, and autoencoders. 
Lastly, our hope is that this primer will inspire readers to explore other niche connections between physics and machine learning -- whether it be the way we can understand ML from physics \textit{or} how ML may help us to understand physics (see for e.g. \cite{alexander2021autodidactic} where just such an idea is explored). 

Before delving into the connection between ML and physics, let us consider some of the features of a computational technique similar to ML that we suspect most will be familiar with, linear regression. The standard picture is that one has a set of data points with the goal to fit some analytical model to the data, perhaps a polynomial function.\footnote{Recall that what makes linear regression \textit{linear} is that the model is linear in the coefficients.} One then optimizes the coefficients of the model such that its deviations from the data are minimized. As we will see later, this is conceptually the basic idea that is underpinning most ML applications. In fact, it is not uncommon to consider linear regression as an elementary example of ML \cite{bishop1995neural,anderson1995introduction}. Not surprisingly, given its simplicity, the effectiveness of regression diminishes with any significant increase in dimensionality, thus the need for ML architectures.

Broadly speaking, machine learning algorithms can typically be placed in one of three main categories: unsupervised, supervised, or reinforcement learning. In this review we will focus on the first two categories but note that reinforcement learning has many applications and is no less important. A key unifying component of all machine learning is the process called \textbf{training}, which is directly related to three classification above, in which data is provided and the algorithm dynamically molds itself to understand this data in the context of how it is trained. This contrasts to more traditional methods which require the user explicitly put in details by hand to achieve a given task. Take a moment to understand what this means. We can use the \textit{same} architecture, often with \textit{no} modification, to attack different problems -- this is simply not possible with traditional approaches.

The first of the three main ML categories, \textbf{supervised} machine learning, is distinguished by its use of \textbf{labeled training data}. For example, if a supervised ML algorithm is fed pictures of dogs and cats for classification, before training, it will likely only get the correct label half of the time. In supervised training we use the correct classification label to inform how we update the model - this is something we will explore in detail later. Performing updates dynamically, the outputs will begin to more closely align with the \textit{known} label for the data. The hope is, that once fully trained, the cat/dog classifier is then able to take in new data and correctly determine whether the images contain a cat or a dog. Supervised learning very popular and has seen many practical application in physics and astronomy, from classifying astronomical objects \cite{Clarke_2020}, improving studies of gravitational lensing data \cite{Godines_2019, Wyrzykowski_2015,Mroz_2020}, to finding new exoplanets \cite{Beaulieu_2006}.

Applications with known data groupings (i.e. with labeled data) are pretty straightforward, but what we if do not know how to group the data in the first place? This is not an uncommon problem; especially in fields like cosmology where data sets are enormous. This idea brings us to the next type of machine learning, unsupervised algorithms. \textbf{Unsupervised} machine learning algorithms execute training with \textbf{unlabeled data}. They are used precisely in cases when we do not have the leisure of knowing what labels to give to our data, and therefore it is up to the algorithm to find relevant correlations among a data set. For example, an unsupervised ML algorithm given a plethora of animal pictures might re-categorize species based on their visible features. Specifically, maybe it would group zebras with striped dogs or spotted cats with spotted dogs. While this algorithm would not have re-created our standard taxonomic system, it might in fact shed light on new properties of the system being examined---e.g. groupings based on fur patterns. The way in which these algorithms learn to cluster data based on their features can also be leveraged in new, powerful ways such as error correction and anomaly detection.

Lastly, \textbf{reinforcement learning} works on a basic principal of positive and negative feedback. For a given task, a reinforcement learning algorithm might be rewarded five points for every correct guess and negative two points for every incorrect guess. It will then utilize the final score to update its algorithm to a higher accuracy. This last main category of machine learning will not be explored further in this primer since supervised and unsupervised learning are far more popular in physics research.\footnote{To learn more about reinforcement learning, see \cite{sutton1998reinforcement}.}

Now that we have provided a bird's-eye view of machine learning, we will now jump straight in. In Section~\ref{HOP} we will introduce Hopfield networks and garner an understanding of its dynamics and training that will serve as a foundation for our understanding of more complicated architectures. We will then move on to Section~\ref{Ising} where we quickly review some basic statistical mechanics and take an in depth look at the Ising model. The deep connection between physics and machine learning is made manifest in Section~\ref{RBM} where we explore Boltzmann machines and discover that they are the Ising model in disguise. Having built our intuition, we then move on to discuss commonly used architectures, feedforward neural networks in Section~\ref{FFNN}, convolutional neural networks in Section \ref{CNN}, and autoencoders in Section~\ref{AE}, where we will again see aspects of physics emerge in their dynamics and training. We then conclude in Section~\ref{Con} where we reflect on the connection between machine learning and physics and discuss the promise for new ideas at the physics/machine learning interface. 

\section{Hopfield Networks} \label{HOP}

To see exactly how some machine learning architectures mimic physics we will begin by reviewing the underpinnings of an architecture known as the \textbf{Hopfield network} \cite{Hopfield2554}.\footnote{For further reading on Hopfield networks, see \cite{hertz_1991,goodfellow2016deep,gerstner2014neuronal}} While Hopfield networks are not particularly useful in practice, their simplistic nature is helpful to understand more general features in machine learning at large. Furthermore, as we will see in this section, the basic structure of this model is eerily close to the Ising model. 

\begin{figure*}[!t]
    \centering
    \includegraphics[width=0.9\linewidth]{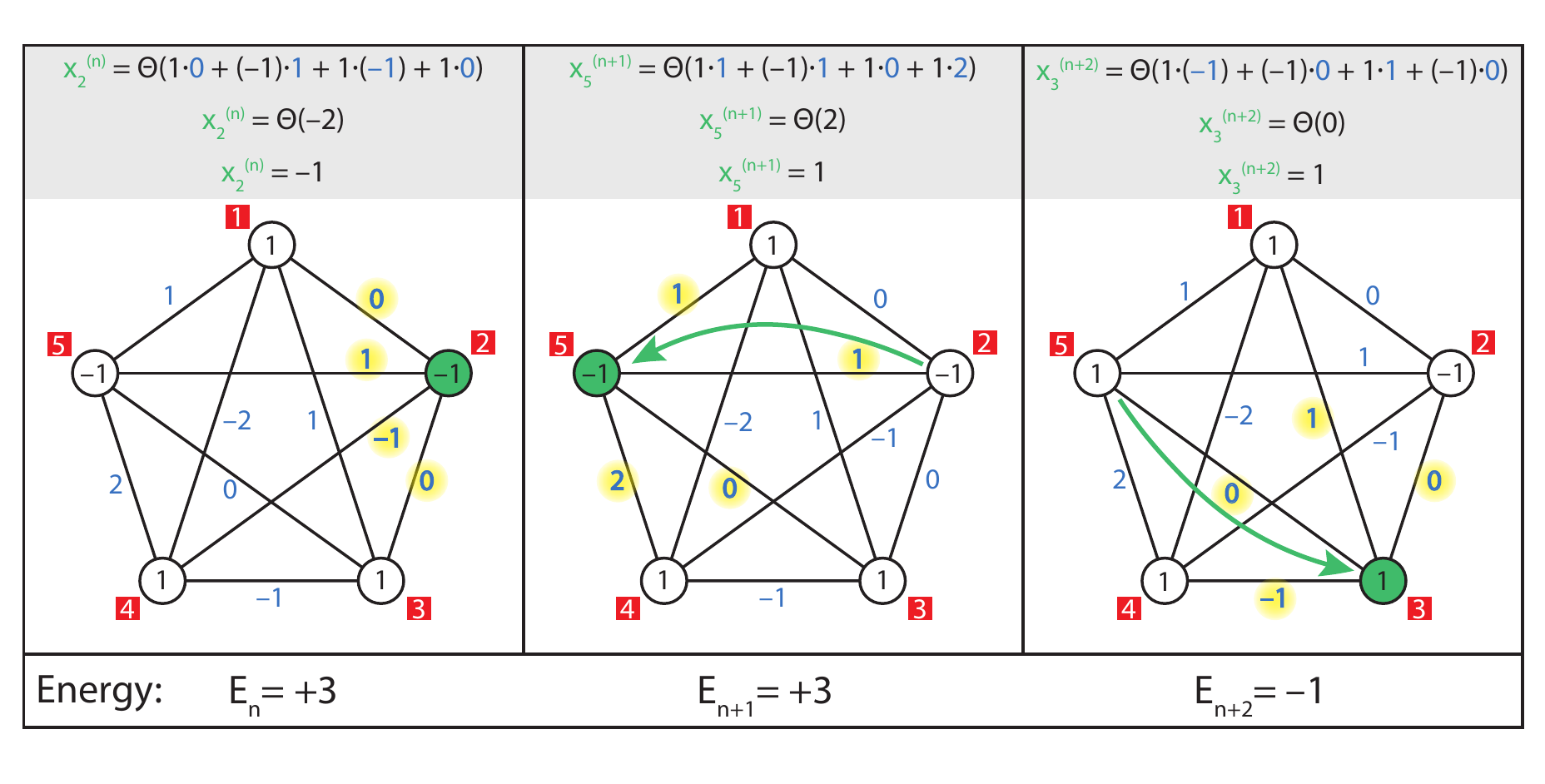}
    \caption{Explicit depiction of the dynamics of the Hopfield network for three updates at times $n$, $n + 1$, and $n + 2$. Red squares are the node label, green nodes represent the current node where we are calculating Eq. \ref{eq:Hopp-dyn}, blue numbers are the connection strength (the $W_{ij}$'s) where we have highlighted relevant weights in calculating the dynamics, and numbers inside nodes represent the state of the network \textit{before} its updated. At the top of each iteration we show the explicit calculation of Eq. \ref{eq:Hopp-dyn} where the spin is updated (if necessary) for the \textit{next} time step. At the bottom we have calculated the energy of the network with Eq. \ref{enhop}. Notice that the energy either stays the \textit{same} or \textit{decreases}.}
    \label{fig:hopfld-dyn}
\end{figure*}

Perhaps the distinguishing features of machine learning algorithms, and which is manifest in Hopfield networks, is the implementation of \textbf{collective computation}. This constitutes many active computations occurring at once within a given algorithm; this is in stark contrast to standard techniques where a set of instructions is tasked to a machine and completed sequentially. The simplest example of collective computation is that of \textbf{associative memory} in a Hopfield network.

The general idea of associative memory is that when an algorithm is presented with some \textit{example} \textbf{X}, the architecture will return some \textit{pattern} \textbf{P}, stored in its \textit{memory}, which most closely resembles \textbf{X} -- this is known as {\it content-addressable memory}. If we assume that our data and patterns take binary input, the simplest architecture which realizes this robustly is based on the McCulloch and Pitts model \cite{mcculloch43a},
\begin{equation}
    x_i^{(n)} = \Theta\left(\sum_j W_{ij} x_j^{(n-1)} - b_i\right),
    \label{MP}
\end{equation}
where $\Theta$ is the Heaviside step function, $x_j^{(n-1)}$ are inputs, $x_i^{(n)}$ is the output, $b_i$ is an activation threshold, and $W_{ij}$ is measure of the relative strength between our output and inputs, and $n$ can be thought of a discrete time. 

\begin{figure*}[!t]
    \centering
    \includegraphics[width=0.8\linewidth]{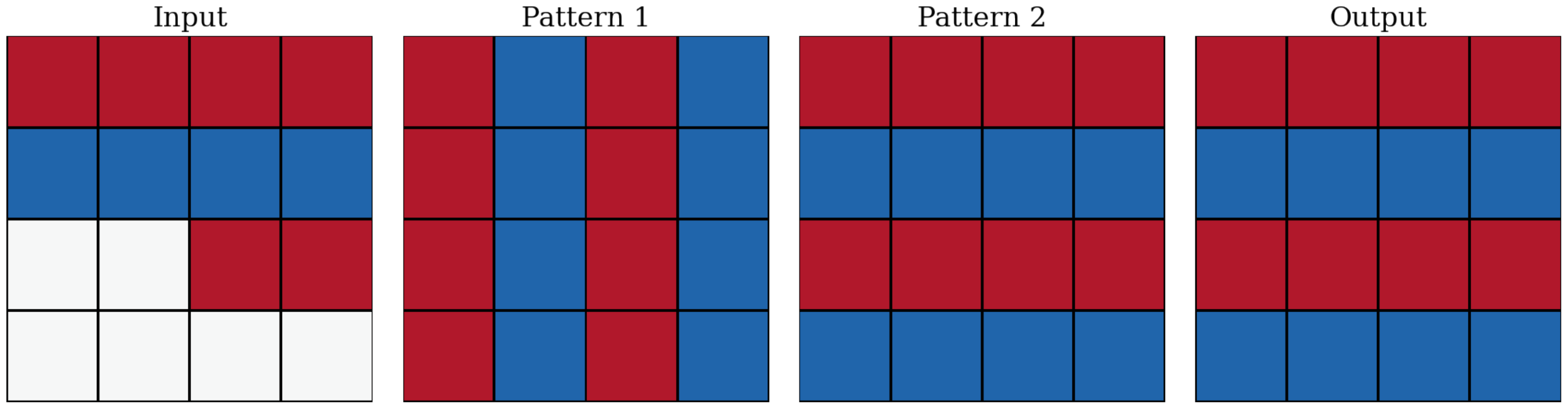}
    \includegraphics[width=0.8\linewidth]{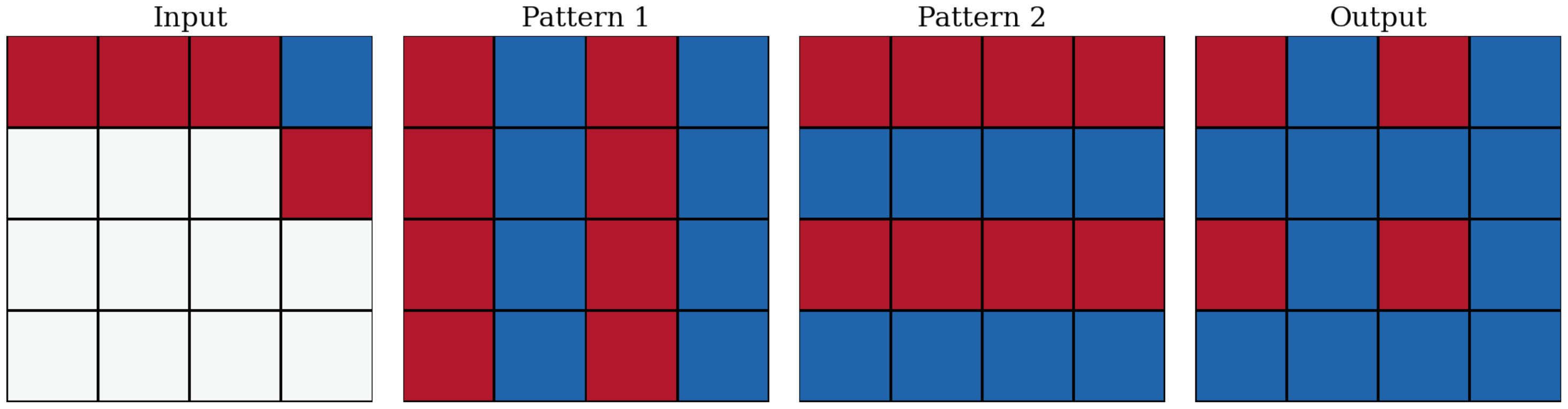}
    \caption{Example of the power of associative memory as realized in a Hopfield network with two stored patterns -- vertical and horizontal stripes. \textit{Top:} Given part of a pattern from its memory as input it correctly reproduces the correct output. \textit{Bottom:} Here we see an extreme example where the architecture fails and becomes \textit{confused}.}
    \label{fig:ass_mem}
\end{figure*}

If we now choose binary units $S_i = \pm 1$ we can rewrite Eq.~\ref{MP} as,
\begin{equation}
    S_i^{(n)} = {\rm sgn}\left(\sum_j W_{ij} S_j^{(n-1)} - b_i\right),
    \label{eq:Hopp-dyn}
\end{equation}
where Eq. \ref{eq:Hopp-dyn} fully dictates the dynamics for the Hopfield network: we move from site to site, randomly, in our architecture and decide the state of nodes based on whether $\sum_j W_{ij} S_j^{(n-1)} \geq b_i$.\footnote{For the remainder of this section we will simply set $b_i = 0$.} We can stop the algorithm once it has converged, i.e. the states no longer change. To aid in understanding, we have depicted the dynamics in an explicit example in Fig. \ref{eq:Hopp-dyn}. Notice one consequence of the threshold feature; it has the benefit of making the architecture more \textit{robust} to errors present in input, i.e. a few incorrect inputs are unlikely to change the final state after repeated application of Eq. \ref{eq:Hopp-dyn}. We present an example in Fig. \ref{fig:ass_mem} where we have passed a \textit{corrupted} version of a pattern stored in a Hopfield network's memory - it has two stored patterns vertical and horizontal stripes. Here the input to the architecture is $4 \times 4$ data where white represent ``corrupted'' pixels that our program must fill in. The robust nature of the network's dynamics are demonstrated in its ability to correctly reproduce the correct pattern even with significant corruption. Not surprisingly, however, if we sufficiently corrupt the data we can ``confuse'' the algorithm as we depict in the second case of Fig. \ref{fig:ass_mem} - one can appreciate that this is truly a extreme example.

How does one choose $W_{ij}$, known  as the weight matrix?\footnote{In this context the weight matrix takes on the form of an adjacency matrix, we will see later that some architectures have a slightly different structure for their weight matrix.} It turns out that it is almost trivial to train a Hopfield network, in fact we can calculate $W_{ij}$ in one step,
\begin{equation}
    W_{ij} = \frac{1}{n} \sum_{\mu}^p P^\mu_i P^\mu_j,
    \label{eq:hop_weight}
\end{equation}
where $\mu$ is an index for $p$ patterns and $n$ is a normalization factor equivalent to the number of neurons. Clearly, then, we can think of the weight matrix in the Hopfield network as a \textit{superposition} of the correlations that encode information for the patterns stored in our memory.
An important feature of the weight matrix, which we had assumed before without stating, is that its is \textit{symmetric}, with $W_{ij} = W_{ji}$,
which is needed for consistent dynamics. This can be understood in terms of the \textit{energy} of the network,
\begin{equation}
    \mathcal{E} = -\frac{1}{2} \sum_{ij} W_{ij}S_iS_j,
    \label{enhop}
\end{equation}
which for the Hopfield network always decreases (or stays the same) when evolved with Eq.~\ref{eq:Hopp-dyn} (see also Fig.~\ref{fig:hopfld-dyn}), reaching a minimum when it has converged on a stored memory.

We now have fully encapsulated the idea of a Hopfield network: \textit{1) Find local energy minimum utilizing symmetric weights and binary threshold units 2) Memories correspond to local energy minima.}

\section{Making Contact with Statistical Mechanics}\label{Ising}

As one should expect, the name \textit{energy} given to Eq.~\ref{enhop} is no accident. Indeed, the astute reader will have likely noticed that Eq.~\ref{enhop} looks surprisingly similar to the Hamiltonian for the Ising model - indeed this was no accident \cite{hertz_1991}. As alluded to earlier, the Ising model is intimately connected to both Hopfield networks and another architecture we will discuss later, Boltzmann machines. In this section we will review the Ising model and develop some tools and intuition that will be useful to us later. However, before we delve into  the Ising model we will briefly review some basic, but important, statistical mechanics.

\subsection*{Review of Relevant Statistical Mechanics}

Recall that entropy of a microstate $\mathcal{S}_1$ is written as $S(\mathcal{S}_1) = k\ln\Omega(\mathcal{S}_1)$ where $k$ is the Boltzmann constant and $\Omega(\mathcal{S}_1)$ is the multiplicity (number of states with the same macroscopic quantity, e.g. energy or volume) of the microstate. For a simple system with binary states, say $\uparrow$ and $\downarrow$, the multiplicity function for $N = N_\uparrow + N_\downarrow$ spins at energy $\mathcal{E}$ is given by $\Omega(N,\mathcal{E}) = N!/(N_\uparrow!N_\downarrow!)$. Taking the large $N$ limit and using the Stirling approximation it is easy to see that the binomial can be approximated as a Gaussian distribution. Furthermore, it is easy to see that the multiplicity function will be sharply peaked - a Gaussian with mean $N$ has width $\sigma = \sqrt{N}$ which will be sharply peaked since $N \gg \sqrt{N}$ in the large $N$ limit. In this sense, when the system has reached thermal equilibrium, it will not deviate too far from its typical state.

It is a natural question to consider the probability that a system is in some state $\mathcal{S}_1$ of energy $\mathcal{E}_1$ or a second state $\mathcal{S}_2$ at energy $\mathcal{E}_2$, both of which are connected to a reservoir $\mathcal{R}$ of energy $\mathcal{E}_R = \mathcal{E}_0 - \mathcal{E}_i$. Here $\mathcal{E}_0$ is the constant, total energy of the entire, isolated system and $i$ is the label for some state. Under the \textit{fundamental assumption} of statistical mechanics, given a sufficient period of time has passed for an isolated system $\mathcal{S}_i + \mathcal{R}$, one is equally likely to be in \textit{any} of the microstates of a system. However, since we specified that we are in some $\mathcal{S}_i$, the number of states accessible to the system is reduced to the multiplicity of $\mathcal{R}$. Under this assumption, clearly the ratio of probabilities to be in a given microstate is equal to ratio of the multiplicities of the reservoirs of the appropriate energy, $\mathcal{E}_0 - \mathcal{E}_i$,
\begin{equation}
    \frac{P(s_1)}{P(s_2)} =              \frac{\Omega(\mathcal{E}_0 - \mathcal{E}_1)}{\Omega(\mathcal{E}_0 - \mathcal{E}_2)}. 
\label{eq:probabilities}
\end{equation}
Since the multiplicity is typically a \textit{very} large number, it is more sensible to rewrite the multiplicity in terms of the entropy,
\begin{equation}
    \frac{P(\mathcal{S}_1)}{P(\mathcal{S}_2)} = e^{\Delta S/k},
    \label{rat}
\end{equation}
where $\Delta S = S(\mathcal{E}_0 - \mathcal{E}_1) - S(\mathcal{E}_0 - \mathcal{E}_2)$. It is not immediately clear how to write the difference in entropies in a nice way. However, if we make the approximation that the energy of the reservoir is significantly larger that the systems we are considering, we can do a Taylor expansion,
\begin{equation}
    S(\mathcal{E}_0 - \mathcal{E}_i) \approx S(\mathcal{E}_0) - \mathcal{E}_i \left(\frac{\partial S_\mathcal{R}}{\partial \mathcal{E}}\right)_{V,N} + \dots,
        \label{taylor}
\end{equation}
with higher order terms vanishing. We can now use the \textit{thermodynamic identity},
\begin{equation}
    T {\rm d}S = {\rm d}\mathcal{E} + P{\rm d}V - \mu {\rm d}N,
\end{equation}
to replace the partial derivative with $1/T$ in Eq. \ref{taylor} since our volume and number of states are fixed (this is what the notation $V,N$ means on the partial derivative term of Eq. \ref{taylor}). Clearly we can now rewrite Eq. \ref{rat} as,
\begin{equation}
    \frac{P(\mathcal{S}_1)}{P(\mathcal{S}_2)} = \frac{e^{-\mathcal{E}_1/T}}{e^{-\mathcal{E}_2/T}},
\end{equation}
which gives us the ratio of probabilities for the system to be in state $\mathcal{S}_1$ at energy $\mathcal{E}_1$ to $\mathcal{S}_2$ at energy $\mathcal{E}_2$. If we are clever, we can now come up with a correct normalization factor such that we can explicitly write down the probability for a single state. Let us consider the sum over the Boltzmann factor for each state $\mathcal{S}$ of the system,
\begin{equation}
    Z = \sum_{{\rm all}~\mathcal{S}} e^{-\mathcal{E}_\mathcal{S}/T},
    \label{part}
\end{equation}
where we now write the probability for some state $\mathcal{S}$ as,
\begin{equation}
    P(\mathcal{S}) = \frac{e^{-\mathcal{E}_\mathcal{S}/T}}{Z}.
\end{equation}
Notice that the probability is well defined,
\begin{equation}
    \sum_{{\rm all}~\mathcal{S}} P(\mathcal{S}) = Z/Z = 1.
\end{equation}
This special quantity, which provides the correct normalization factor to define probabilities, is called the \textbf{partition function} and it is an understatement to say that it is the backbone of statistical mechanics. This stems from the fact that there are \textit{many} quantities that can be calculated directly from the partition function. One that we will find useful in the next subsection is that the average value for some quantity $X_\mathcal{S}$ is given by,
\begin{equation}
    \left< X \right> = \frac{1}{Z} \sum_{{\rm all}~\mathcal{S}} X_\mathcal{S} e^{-\mathcal{E}_\mathcal{S}/T}.
    \label{average}
\end{equation}

This concludes our review of relevant statistical mechanics. We will see in several places, particularly in the contexts of Boltzmann machines in Section \ref{RBM}, where many of the basic principles of statistical mechanics we have just discussed emerge in the context of machine learning.

\subsection*{The Ising Model}

The \textbf{Ising Model} is a model of ferromagnetism where atoms in a solid (lattice) can be either spin-up or spin-down ~\cite{mccoy2014two}. The system has a temperature that can increase or decrease, and thus can alter the energy of the system as a whole. If two neighboring lattice sites have spins in the same direction they are in a lower energy state, while neighboring sites with spins of opposite directions are in a higher energy state. This tendency for spin alignment can become overpowered when the lattice is at a higher temperature -- thermal fluctuations introduce randomness into lattice -- but as the lattice is cooled, it will undergo a second order phase transition and the spins will begin to align. In addition to temperature, one can also introduce an external magnetic field $B$ which can cause the system to undergo a first order phase transition. 

The system has the following Hamiltonian,
\begin{equation}
H = - J\sum_{<i,j>}\sigma_i\sigma_j - \sum_i B\sigma_i,
\label{eq:ising_energy}
\end{equation}
where we sum over the spins of neighboring sites which can take on values of -1 and 1. $J$ represents the interaction strength between any two sites in the lattice and $B$ represents the external magnetic field applied to the system. Note that the explicit negative sign in Eq. \ref{eq:ising_energy} forces parallel (anti-parallel) spins to correspond to a low (high) energy state.  

While the Ising model can be solved exactly in 1- and 2-dimensions, solutions in higher dimensions are less tractable. One approach to get around this issue is the application of mean field theory. The basic principle is to isolate spins in the lattice by replacing interactions with neighbors by an \textit{effective} field experienced by individual spins,
\begin{equation}
    \left<B_{i}\right>  =  \frac{J}{N}\sum_{j}\left<\sigma_j\right> - B,
\end{equation}
where $\left<\sigma_j\right>$ is the average spin in the lattice. We can rewrite this in terms of the order parameter $m$ (i.e. magnetization) and coordination number $q$ (number of neighbors),
\begin{equation}
    \left<B_{i}\right>  = J q m - B.
\end{equation}

\begin{figure}[t]
    \centering
    \includegraphics[width=0.9\linewidth]{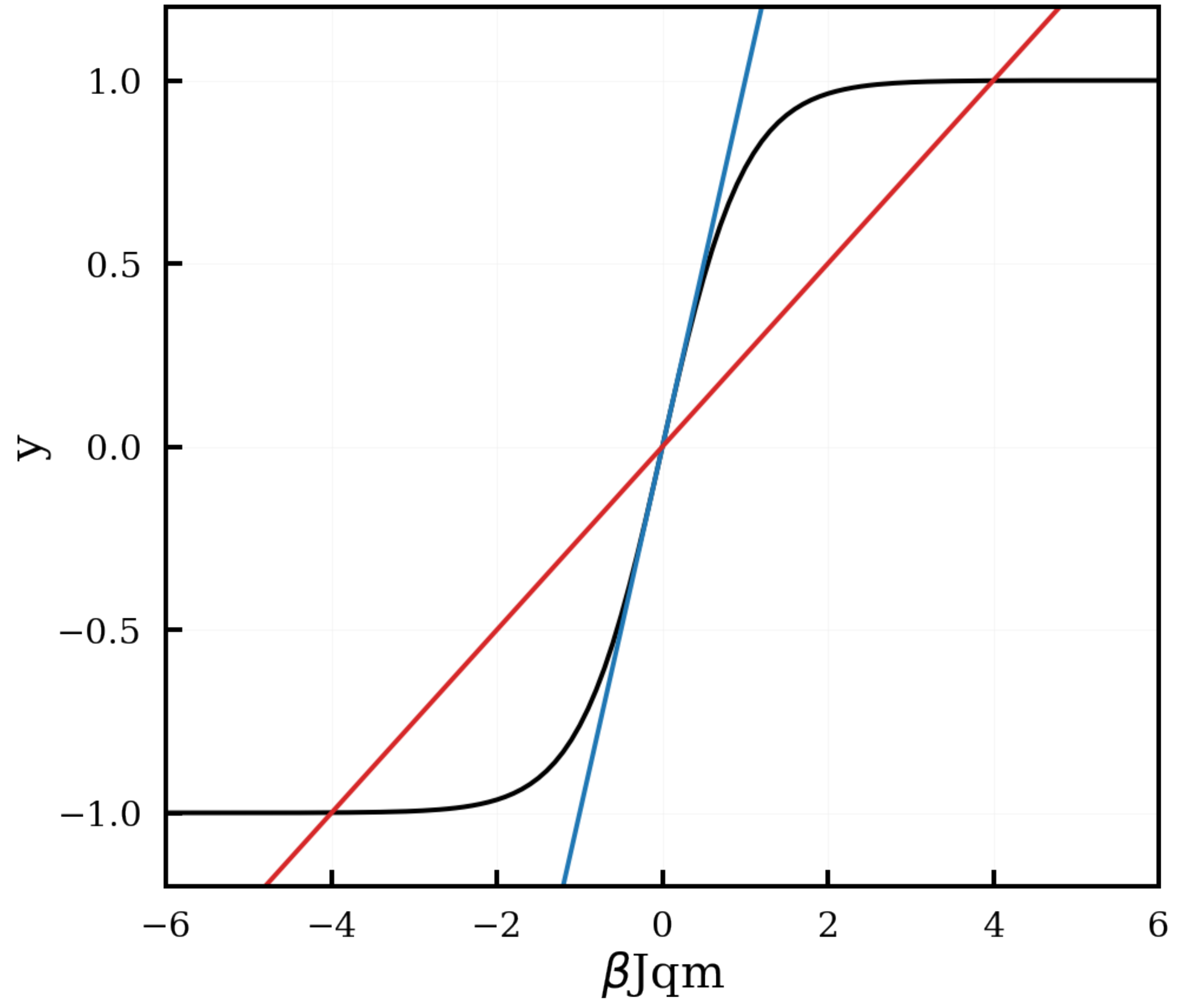} \caption{We can visualize the solutions for Eq. \ref{mag} at different temperatures. The blue curve corresponds to a high temperature state with a single net zero spin solution and the orange curve is a low temperature with two solutions corresponding to the all spin up/down states.}
    \label{fig:mft}
\end{figure}
Since the goal of mean field theory is to have an isolated spin with our new effective field, we can begin by solving the single particle problem. In this case it is simple to write down the average spin, as it is only two terms, using Eqs. \ref{part} and \ref{average},
\begin{equation}
    \left< \sigma_i \right> = \frac{e^{-\beta \left<B_{i}\right>}}{e^{-\beta \left<B_{i}\right>} + e^{\beta \left<B_{i}\right>}} - \frac{e^{\beta \left<B_{i}\right>}}{e^{-\beta \left<B_{i}\right>} + e^{\beta \left<B_{i}\right>}},
\end{equation}
which we can conveniently rewrite as,
\begin{equation}
    m = {\rm tanh}\left(\beta J q m - \beta B\right).
    \label{mag}
\end{equation}
If we assume that there is no background magnetic field $B$, it is relatively straightforward to solve for the order parameter. We show a couple solutions in Fig. \ref{fig:mft}. For this {\it graphical} solution the x-axis is ${\beta}Jqm$ and we have graphed $y=\tanh(x)$, and $y = \frac{x}{{\beta}Jq}=m$. The two different linear lines correspond to the different values of $\beta$: blue is above the phase transition (only a single solution) and red is below the phase transition (two solutions). At the intersection points, we find where $m = \tanh({\beta}Jqm)$, or $\frac{x}{{\beta}Jqm}=\tanh(x)$.  

An interesting thing to point out is that for binary units of spin 0, 1, (instead of -1 and 1) this corresponds to a function of the form
\begin{equation}
    f(x) = \frac{1}{1 + e^{-x}},
    \label{sig_def}
\end{equation}
which is the well known \textit{sigmoid} activation function, an integral part of neural networks -- we will discuss activation functions in more detail when we discuss feedforward neural networks. It is certainly interesting that these activation functions both naturally arise from modeling the average magnetization of the Ising model. Indeed, it may not be too far of a stretch to claim that a trained neural network, which we will also discuss in detail later, exists in a something analogous to a \textit{critical} state. Thereby, it has the flexibility, for example, to respond to inputs and classify them in one of two states.\footnote{See \cite{Roberts:2021fes} where the notion of criticality in neural networks is explored in detail.}

\section{(Restricted) Boltzmann Machines}\label{RBM}

\begin{figure}[!t]
    \centering
    \includegraphics[width=0.49\columnwidth]{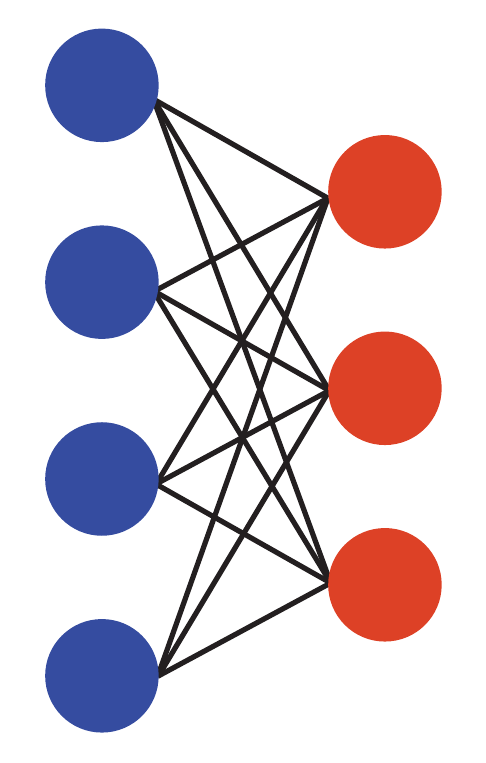}
    \caption{The structure of an RBM: the blue circles represent the visible layer while the orange circles represent the hidden layer.}
    \label{fig:rbm}
\end{figure}

Hopfield networks are not the only algorithms that inherited the structure of the Ising model, another closely related architecture is the \textbf{Boltzmann machine} (BM) \cite{hertz_1991}. This architecture is a generative, stochastic, fully connected, artificial neural network with hidden units: generative meaning it can be used to ``create'' data, stochastic meaning that it relies on a learned probability distribution, and hidden units meaning there are nodes that are not used as inputs/outputs (these nodes are called visible). Put another way, a BM is just a stochastic Hopfield network with hidden units. This means that the defining difference is that it has been trained to learn a probability distribution - i.e. it is not deterministic. BMs can be implemented as either an unsupervised or supervised algorithm. We will restrict our attention to the former, but most of our discussion translates directly to the supervised case. A common version of the BM, which like Hopfield networks is a fully connected graph, is the {\bf restricted Boltzmann machine} (RBM) \cite{6302931} where one \textit{restricts} connections in the architecture to eliminate interactions between nodes of the same group, hidden or visible, such that the architecture forms a bipartite graph - see Fig.~\ref{fig:rbm}.\footnote{For more information on RBMs, see \cite{goodfellow2016deep,zocca2017python}} 

\subsection*{RBM Structure}
From now on we restrict our discussion to RBMs. As we mentioned previously, the RBM can be though of as two sets of nodes represented as vectors. One set is the \textbf{visible layer} and the second is the \textbf{hidden layer}. The two layers are connected by a weight matrix $\mathbf{W}$. Just like in the Hopfield network (and Ising model) we can compute a scalar value that represents the state of the system,
\begin{equation}
    \mathcal{E}_{\rm RBM} = - \mathbf{h}^{\rm T} \mathbf{W} \mathbf{v} - \mathbf{h}\mathbf{b}^{\rm T}_{\mathbf{h}} - \mathbf{v} \mathbf{b}^{\rm T}_{\mathbf{v}}.
    \label{eq:rbm_energy}
\end{equation}
which we see mirrors the Ising model Hamiltonian (Eq.~\ref{eq:ising_energy}).
Here $\mathbf{v}$ and $\mathbf{h}$ correspond to the visible layer and the hidden layer, respectively, $\mathbf{W}$ corresponds to the weight matrix, and $\mathbf{b_v}$, $\mathbf{b_h}$ correspond to biases for their respective layer. One thing to note before moving on is that the weight matrix for an RBM, when treating the visible and hidden layers as vectors, is not the same weight matrix as the more general Boltzmann machine and Hopfield network - specifically the weight matrix you use in practice, though one can always write the RBM weight matrix as an adjacency matrix.\footnote{We emphasize this difference by using $W_{ij}$ and $\mathbf{W}$. This is actually why there is a factor of 2 difference between Eqs.~\ref{enhop} and~\ref{eq:rbm_energy}.} We will discuss some of these difference in more detail in our discussion of feedforward neural networks where the weight matrix takes a similar form to RBMs (though its actually still not quite the same). Now looking at Eq. \ref{eq:rbm_energy}, in analogy to the Ising model Hamiltonian (Eq. \ref{eq:ising_energy}), the first term corresponds to nearest neighbor interactions and the last two bais terms are analogous to the magnetic field terms. One important difference is that the weight matrix for a RBM does not need to be restricted to binary values - for the Ising model spins are either neighbors or they or not.

Given the similarities to the Ising model that we have just demonstrated, it is probably not that surprising that the probability to be in a given configuration $\alpha$ of the \textit{visible} layer is given by,\footnote{Notice that this is typically what one would be concerned about, i.e. this will correspond to the networks representation of your training data.}
\begin{equation}
    P_{\alpha} = \frac{1}{Z} \sum_{\alpha} e^{-\mathcal{E}_{\rm RBM,\alpha}/T},
\end{equation}
where we sum over all possible configurations of the \textit{hidden} layer in such a way that we account for all states of the whole system with the \textit{visible} layer in the state $\alpha$. The normalization is of course just the partition function for the system. As we will see later, it is the need to sample this partition function to update weights which makes training computationally expensive, i.e. one must sample exponentially many terms to reach thermal equilibrium. 

\subsection*{RBM Dynamics}

With a solid understanding of the structure of RBMs, let us now consider its dynamics. We first begin by placing our data onto the visible layer, $\mathbf{v}$, which we can imagine for binary units looks something like $(1,0,0,\dots)^T$. Given our choice of structure for the weight matrix and the fact that $\mathbf{v}$ and $\mathbf{h}$ are vectors, we can go from one layer to next as,
\begin{equation}
    \mathbf{P_h} = \sigma(\mathbf{W v}),
    \label{eq:prob}
\end{equation}
where $\sigma$ is an activation function, here a sigmoid function (Eq. \ref{sig_def}), and $\mathbf{P_h}$ is now interpreted as a distribution of probabilities over the hidden layer. We can interpret these probabilities as giving a measure of how likely nodes are to be \textit{on} or \textit{off}.

\begin{figure*}[t]
    \centering
    \includegraphics[width=0.65\linewidth]{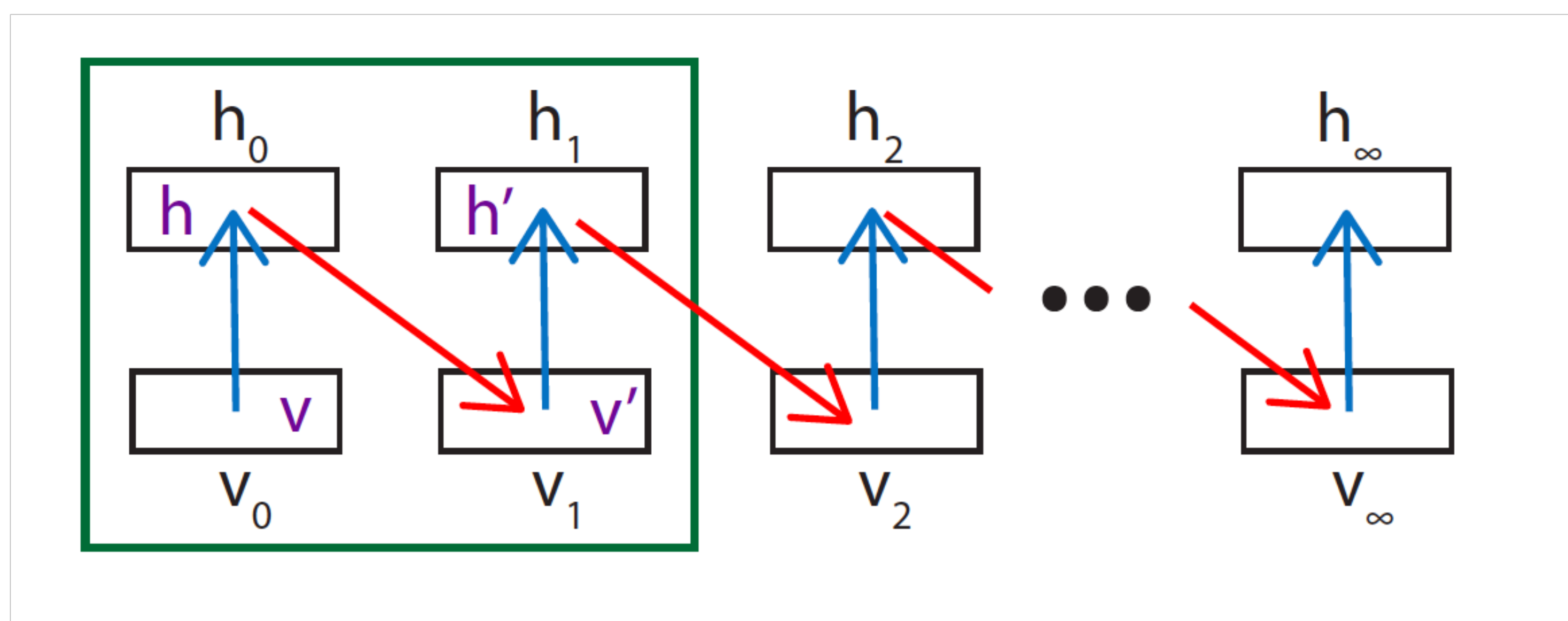}
    \caption{Depiction of maximum likelihood learning. One iterates back and forth from the visible to the hidden layer, all the while using Gibbs sampling, until the system is thermalized, here represented by $v_\infty, h_\infty$. Contrastive divergence on the other hand only considers the system after one iteration, i.e. the green box in the figure. Naturally, since one must calculate exponentially many terms for maximum likelihood learning, contrastive divergence is a much quicker algorithm.}
    \label{fig:ml}
\end{figure*}

We then decide the state for the layer using {\bf Gibbs sampling}: we generate a random number from a uniform distribution between zero and one for each node and if this random value is greater(less) than the corresponding element from $\mathbf{P_h}$ we turn that node \textit{on}(\textit{off}). Let us now consider a concrete example -- if $\mathbf{P_h}$ was found to have the following values for the \textit{first} pass to the hidden layer,
\begin{align}
    \mathbf{P_h}  &= \begin{pmatrix}
           0.6001 \\
           0.8452 \\
           0.5789 \\
           0.9132 \\
           0.1159
         \end{pmatrix},
\end{align}
and one generated the following random values
\begin{align}
    \mathbf{R}  &= \begin{pmatrix}
           0.9777 \\
           0.1698 \\
           0.1054 \\
           0.7568 \\
           0.2337
         \end{pmatrix},
\end{align}
using the condition that $R_i>P_i$, we input 1 into the corresponding element of a new vector $\mathbf{h}$ and if $R_i<P_i$, we input 0 into the corresponding element of $\mathbf{h}$. Performing this operation on our example we have found the hidden layer,
\begin{align}
    \mathbf{h}  &= \begin{pmatrix}
           1 \\
           0 \\
           0 \\
           0 \\
           1
         \end{pmatrix}.
\end{align}
Going \textit{back} to the visible layer there is one subtlety, when going from the hidden layer back to the input layer we must of course use the \textit{transpose} of the weight matrix to construct our probability distribution of the inputs,
\begin{equation}
    \mathbf{P}'_\mathbf{v} = \sigma(\mathbf{W}^T\mathbf{h}),
\end{equation}
which we can then sample from to yield our updated input layer $\mathbf{v}'$; note that we use the $'$ to emphasize that $\mathbf{v} \neq \mathbf{v}'$. Iterating this process repeatedly results in effectively sampling the partition function for the system, for this reason it is generally referred to as \textbf{thermalization}. In the next step we will discuss the means to train a Boltzmann machine. We will find that the goal is to find the set of weights such that when when brought to thermal equilibrium (using the above process) yields a distribution that approaches that used for training. In other words, we train the weights in such a way that the data set is the minimum energy state of system. 

\subsection*{Training an RBM}

As we just alluded to, the goal in training a Boltzmann machine is to learn a distribution over the input nodes that matches that for the training set. If our data set follows a distribution $\mathbf{Q_\alpha}$ and our nodes have distribution $\mathbf{P_\alpha}$ we quantify the difference between the distributions via their relative entropy,
\begin{equation}
    S = \sum_\alpha Q_\alpha \log{\frac{Q_\alpha}{P_\alpha}}.
\end{equation}
Since our goal is to minimize the separation between the distributions, this equates to minimizing the relative entropy. It is then natural to calculate the gradient of $S$ with respect to the weight matrix,
\begin{equation}
    \Delta \mathbf{W} = \frac{\partial S}{\partial \mathbf{
    W}} = - \left< \mathbf{h v} \right>_{\rm clamped} + \left< \mathbf{h v} \right>_{\rm free},
    \label{weight_up}
\end{equation}
where $\left< \mathbf{h v} \right>_{\rm clamped}$ represents the average at thermal equilibrium with the data vector \textit{clamped}, we don't allow nodes of $\mathbf{v}$ to change, and $\left< \mathbf{h v} \right>_{\rm free}$ is the average at thermal equilibrium where all spins are allowed to change. Hopefully it is now clear what was meant earlier by training taking a significantly long time; for every weight update we must reach thermal equilibrium to calculate terms in Eq. \ref{weight_up}. In practice this is realized as the back and forth procedure discussed earlier that brings the system to thermal equilibrium. This general procedure for calculating weight updates is referred to as {\bf maximum likelihood learning} which we depict schematically in Fig.~\ref{fig:ml}. 

While this is the most robust method for training a RBM, the method of {\bf contrastive divergence}, proposed by Hinton as a cheap alternative, is more widely used~\cite{10.1162/089976602760128018}.\footnote{Simulated annealing can also be used for training RBMs.} Indeed, the latter is more precise but computationally expensive as one requires the calculation of an exponential number of terms to get the system to reach true equilibrium. While contrastive divergence has been shown to work quite well empirically, it necessarily is biased by only considering a few Markov chains. Indeed, it was shown that the fixed points for contrastive divergence differ from those of maximum likelihood learning but the bias for most scenarios is small \cite{inproceedings}. 

Concretely, contrastive divergence terminates this back and forth procedure after a single full sweep, see the green box in Fig. \ref{fig:ml}, which naturally leads to faster weight updates. The standard notation to denote this update is thus given by, based on labels in Fig \ref{fig:ml},
\begin{equation}
    \Delta {\bf W} = - \mathbf{h v} + \mathbf{h' v'}.
    \label{weight_ud_mll}
\end{equation}
We can then update $\mathbf{W}$ by adding ${\Delta}\mathbf{W}$, scaled by a factor called the learning rate, $\eta$,
\begin{equation}
    \mathbf{W}_{\rm new} = \mathbf{W} + \eta{\Delta}\mathbf{W}.
\end{equation}
Another useful metric to keep track of the performance for a RBM is to calculate the \textbf{mean squared error} (MSE) between the input data, $\mathbf{v}$, and the reconstructed visible layer, $\mathbf{v}'$,
\begin{equation}
    \rm{MSE} = \frac{1}{n} (\mathbf{v} - \mathbf{v'})^2,
\end{equation}
where $n$ is the dimension of the input data. It is easy to see from this expression that the MSE will decrease as the RBM becomes more competent in reconstructing the original data from its learned, internal, compressed representation. This speaks to interpreting what is actually happening in the RBM; one can think of nodes of the hidden layer corresponding to \textit{features} of the inputs  while \textit{correlations between features} are encoded by the weights. To understand this in some detail we will now consider a hypothetical application of RBMs.

\subsubsection*{Example Application of a RBM}

A classic application of BMs/RBMs include the prediction of movie ratings. Imagine a data set of 50,000 movies and binary reviews (i.e. like or dislike) by 1,000 people -- with the inevitability that for a given person only a fraction of these movies have been seen. We could of course represent each persons' movie watching history with an input vector where each entry has either a -1 for disliked movies, a +1 for liked movies, and a 0 for unseen movies. The RBM will learn a  compressed, higher level representation of the data set in the hidden nodes during training and relevant correlations between the input layer as represented by the weights. A qualitative flavor of this could be that some nodes are trained to identify genres of movies and others identify movies with specific actors. Correlations between these compressed representations allow the  architecture to make predictions for movies that a user has not seen, i.e. based on correlations it learned from other users. It is this sense that makes the architecture generative. Notice how this is exactly analogous to the behavior of corrupted input and associative memory in Hopfield networks. 

Now that we have firmly established the intimate connection between energy based machine learning algorithms and physics, we would now like to move on to describing some of the most popular machine learning architectures.

\section{Feedforward Neural Networks}\label{FFNN}

By far the most popular architectures in machine learning are based on {\bf feedforward neural networks}, which we will sometimes refer to as a neural network (NN) for simplicity. At a high level, one can thing of a neural network as map from a high dimensional space to a smaller space. In this sense we can think of it as a function - given an image of a cat it returns the answer \textit{yes} or \textit{no}. A simple example is a single \textbf{layer} linear neural network,
\begin{equation}
    \mathbf{y} = \mathbf{W} \mathbf{x} + \mathbf{b},
    \label{eq:fcNN}
\end{equation}
where $\mathbf{x}$ is an input to the architecture (say an image), $\mathbf{W}$ is the weight matrix, $\mathbf{b}$ the bias, and $\mathbf{y}$ is the output. Notice the similarity in the structure to that of the perceptron, Eq. \ref{MP}. Indeed, we can think of a neural network as being a higher dimensional generalization of a perceptron. While it appears that the structure is also very similar to a RBM, i.e. they are both bipartite graphs, we will see later the dynamics are actually quite different! This simple feature, that NNs form a bipartite graph, is very convenient computationally.  Notice that with $\mathbf{W}$ represented as a matrix and $\mathbf{y,~x,~\&~b}$ all vectors we can utilize the convenience of matrix multiplication were we are easily able to transition from one layer to the next, in the same way that we could for RBMs.

\begin{figure*}[t]
    \centering
    \includegraphics[width=\linewidth]{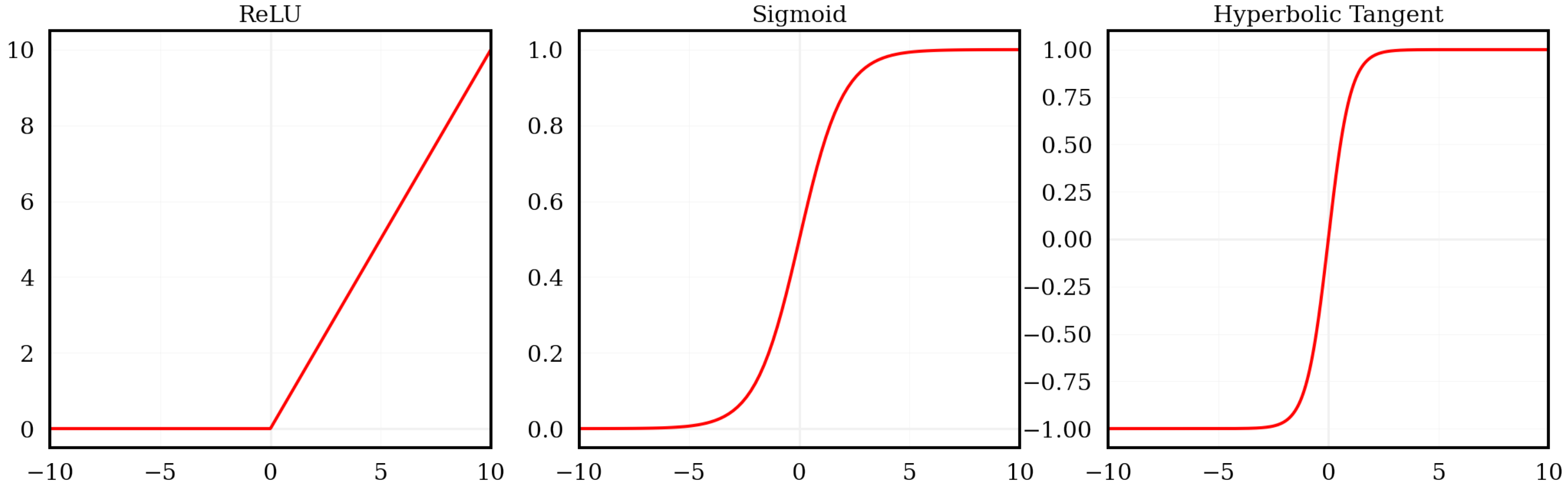}
    \caption{From \textit{left to right}, the ReLU, sigmoid, and tanh activation functions.}
    \label{fig:act}
\end{figure*}

Let us now cover in detail the dynamics of a single layer NN. Consider some input that is multiplied by the weight matrix, $\mathbf{W}$, which in this context can be thought of as a weighted averaging operator or as rotation matrix from the input to the feature space. We then offset this by the bias which amounts to shifting a threshold, or, more simplistically, it is just the higher dimensional generalization of setting a y-intercept for a line when working in $\mathbb{R}^2$. If we now undress Eq.~\ref{eq:fcNN} we have, where $\mathbf{W}$ is an $M \times N$ matrix and $M$, $N$ are the output and input dimensions respectively,
\begin{equation}
\begin{pmatrix}
y_{\bar1}&\\
y_{\bar2}&\\
\vdots&\\
y_{\bar m}\\
\end{pmatrix}
=
\begin{pmatrix}
W_{\bar11} & W_{\bar12} & \dots & W_{\bar1n} \\
W_{\bar21} & W_{\bar22} & \dots & W_{\bar2n} \\
\vdots & \vdots & \ddots & \vdots\\
W_{\bar m1} & W_{\bar m2} & \dots & W_{\bar mn} \\
\end{pmatrix}
\begin{pmatrix}
x_1&\\
x_2&\\
\vdots&\\
x_n\\
\end{pmatrix}
+
\begin{pmatrix}
b_{\bar{1}}&\\
b_{\bar{2}}&\\
\vdots&\\
b_{\bar{m}}\\
\end{pmatrix},
\label{exp}
\end{equation}
where we have placed a bar over indices corresponding to the \textit{output} layer to distinguish them from the input layer. Consider what this means by noticing that $W_{\bar{2} 1}$ is the weight between $y_{\bar{2}}$ and $x_1$ and $W_{\bar{1} 2}$ is between $y_{\bar{1}}$ and $x_{2}$. Notice that the weight matrix here has a structure different from that generally used for BMs and Hopfield Networks. In those contexts the weight matrix is generally taken to be in the form of an adjacency matrix; i.e. there $W_{12}, W_{21}$ describe connections between the \textit{same} nodes, but notice they are \textit{directed}. Recall that we made the argument that the weight matrix for the Hopfield network and BM should be symmetric such that we could define a consistent energy for the architecture. Naively one sees that $\mathbf{W}$ as defined for a NN is \textit{not} symmetric in Eq. \ref{exp}. Indeed, even if rewritten as a proper adjacency matrix their weights are still \textit{not} symmetric since these architectures are \textit{feedforward}, i.e. there is no path $y_{\bar{m}} \rightarrow x_n$. We urge the reader to make sure that they understand this subtle difference. Indeed, while the structure of feedforward neural networks (Eq. \ref{eq:fcNN}) and Boltzmann machines (Eq. \ref{eq:prob}) look similar, they are \textit{not} the same thing.

Single layer neural networks (Eq. \ref{eq:fcNN}) are not very useful in practice as one typically constructs architectures composed of multiple layers. Of course, each layer has its own weight matrix and bias vector where inputs are taken as the outputs of the previous layer -- excluding, of course, the first such layer. There is one other practical issue with Eq. \ref{eq:fcNN}. Indeed, an astute reader will have realized that it is not particularly useful since it is \textit{linear}. Even the composition of multiple layers of this form is still linear; consider the following two-layer NN, with zero bias, for simplicity,
\begin{equation}
    \mathbf{y}_2 = \mathbf{W}_2 \mathbf{y}_1 = \mathbf{W}_2 (\mathbf{W}_1 \mathbf{x}_1) = \hat{\mathbf{W}}_2 \mathbf{x}_1,
\end{equation}
where $\hat{\mathbf{W}}_2 = \mathbf{W}_2 \mathbf{W}_1$. So the composition of multiple layers looks like a single layer NN with weight matrix $\hat{\mathbf{W}}_2$. It is now easy to see that no matter the depth of our architecture our mapping $\mathbf{y}(\mathbf{x})$ will always be linear, which is not useful for problems of any interest. The solution is to introduce a non-linearity in our mapping between layers. This is most readily done with a \textit{nonlinear} activation function, $\sigma(z)$, such that we now have,
\begin{equation}
    \mathbf{y} = \sigma(\mathbf{W} \mathbf{x} + \mathbf{b}).
    \label{eq:fcNN_act}
\end{equation}
Some common activation functions are ReLU (Rectified Linear Unit),
\begin{equation}
     \sigma(\mathbf{z}) = \left\{
\begin{array}{ll}
      \mathbf{z} ~~~ \mathbf{z} \geq 0 \\
      0 ~~~ \mathbf{z} < 0 \\
\end{array}, 
\right. \\
\label{relu_act}
\end{equation}
sigmoid,
\begin{equation}
    \sigma(\mathbf{z}) = \frac{1}{1 + e^{-\mathbf{z}}},
\end{equation}
and hyperbolic tangent,
\begin{equation}
    \sigma(\mathbf{z}) = \tanh(\mathbf{z}),
\end{equation}
which we depict in Fig. \ref{fig:act}.

With the non-linearity induced by the activation function its then sensible to compose multiple layers. In practice, one typically does this by composing layers that also progressively decreases the dimension from the previous layer. With exceptions,\footnote{A common example is if the complexity of your architecture exceeds that of your data set. In this case, one is prone to overfitting. } performance is improved by increasing the depth of neural networks and, while there isn't a strict definition, architectures with more than 5 layers constitute \textbf{deep learning}. These intermediate compressed representations, the so called hidden layers, represent higher level \textbf{features} of the input. Indeed, this is analogous to our discussion with the hidden layer of RBMs where the nodes learn a representation of higher dimensional features from the data set. As a qualitative example, perhaps some nodes of a trained cat-dog classifier become stimulated when there is a preponderance of orange in image that leads to the classification of ``cat.'' A real-world, high level feature which it may have learned during its training is to correlate the color orange with cats. 

While we will not delve too far into the details, we think it worthwhile to point out to the interested reader the connection between learned features in NNs and the renormalization group (RG). Indeed, for a trained architecture, there is a clear sense in which going from layer to layer of a NN is representative of a RG flow. Relevant degrees of freedom propagate and irrelevant degrees of freedom are integrated out under the mapping learned through training. For those not immediately convinced, consider again the simple example of a classifier trained to identify cats and dogs; such an architecture is able to compress an image down to a binary classification label. Given such architectures are easy to train to near perfect classification, one must appreciate that such architectures must be implementing an RG-like procedure, i.e. they clearly learn the relevant degree of freedom - given an image of a dog, it informs us that there is a dog present.\footnote{For those interested in the ML connection to the RG we suggest \cite{Roberts:2021fes,2014arXiv1410.3831M}.}

\subsection*{Training Neural Networks}

So far in our discussion of NNs we have considered architectures which have already been trained. Naturally, of course, one must also train their feedforward neural network for a given data set and desired application.  Similar to the RBM, we must consider the gradient of a loss function with respect to the architecture's weights. This process is referred to as \textbf{backpropagation} and differs from maximum likelihood learning and contrastive divergence in RBMs, most notably in that the dynamics are deterministic and the choice of loss function \cite{back-propagation}. A standard choice of loss function is the mean squared error,
\begin{equation}
    \mathcal{L} = \frac{1}{n}\left(\mathbf{y}_{\rm out} - \mathbf{y}_{\rm known}\right)^2,
    \label{mse_loss}
\end{equation}
where $\mathbf{y}_{\rm out}$ is the output of the layer, $\mathbf{y}_{\rm known}$ is the known~(desired) output, and $n$ represents the size of the output. 

To illustrate the process of backpropagation, will work out an explicit example. Consider a single layer feedforward neural network with ReLU activation function, Eq. \ref{relu_act},
\begin{equation}
    \mathbf{y} = \sigma(\mathbf{z}),
    \label{act_y}
\end{equation}
where we define $\mathbf{z}$ as the layer without activation,
\begin{equation}
    z = \mathbf{W} \mathbf{x} + \mathbf{b}.
\end{equation}
Now consider a MSE loss function, Eq. \ref{mse_loss}, such that we calculate the gradient as,
\begin{equation}
    \Delta{\mathbf{W}} = \frac{\partial{\mathcal{L}}}{\partial{\mathbf{W}}} = \frac{\partial{\mathcal{L}}}{\partial{\mathbf{y}_{\rm out}}}\frac{\partial{\mathbf{y}_{\rm out}}}{\partial\sigma}\frac{\partial\sigma}{\partial{\mathbf{z}}}\frac{\partial{\mathbf{z}}}{\partial{\mathbf{W}}},
    \label{Delta_W}
\end{equation}
where we have used the \textit{chain rule} to get the last equality. Let us now consider the contribution from each derivative on the r.h.s. of the last equation. We begin by taking the derivative of the loss function with respect to our output,
\begin{equation}
    \frac{\partial \mathcal{L}}{\partial \mathbf{y}_{\rm out}} = \frac{2}{n} \left( \mathbf{y}_{\rm out} - \mathbf{y}_{\rm known}\right).
\end{equation}
The second partial derivative is just unity, i.e. see Eq.~\ref{act_y},
\begin{equation}
    \frac{\partial{\mathbf{y}_{\rm out}}}{\partial\sigma} = 1.
\end{equation}
Next we consider the derivative of the ReLU activation function, Eq. \ref{relu_act},
\begin{equation}
        \frac{\partial\sigma}{\partial{\mathbf{z}}} = \left\{
\begin{array}{ll}
      1 ~~~ \mathbf{z} \geq 0 \\
      0 ~~~ \mathbf{z} < 0 \\
\end{array}. 
\right. \\
\end{equation}
We can now calculate the final partial derivative from Eq.~\ref{Delta_W},\footnote{Where we use the following result for derivatives of vector quantities with respect to matrices, ${\rm d} (\mathbf{W x}) / {\rm d} \mathbf{W} = \mathbf{x}^{\rm T}$.}
\begin{equation}
    \frac{\partial{\mathbf{z}}}{\partial{\mathbf{W}}} = \frac{\partial}{\partial{\mathbf{W}}} \left( \mathbf{W} \mathbf{x} + \mathbf{b} \right) = \mathbf{x}^{\rm T}.
\end{equation}
Now we can bring everything together for $\Delta W$,
\begin{equation}
    \Delta \mathbf{W} = \left\{
\begin{array}{ll}
       \frac{2}{n}\left( \mathbf{y}_{\rm out} - \mathbf{y}_{\rm known}\right)\mathbf{x}^{\rm T}  ~~~ \mathbf{z} \geq 0 \\
      0 ~~~~~~~~~~~~~~~~~~~~~~~~~~~~ \mathbf{z} < 0 \\
\end{array}. 
\right. \\
\end{equation}
This means that we \textit{do not} update weights for elements where $\mathbf{z} = \mathbf{W}\mathbf{x} + \mathbf{b}$ are negative. This is the so called \textit{vanishing gradient problem} which can make training with ReLU difficult.\footnote{This problem can be addressed by modifying ReLU to take small but non-zero values for $z < 0$; this is called leaky ReLU.} With our gradient now in hand we can update our weights such that we take a step in the loss landscape, minimizing our loss,
\begin{equation}
    \mathbf{W}^{\rm new} = \mathbf{W} + \eta\Delta{\mathbf{W}},
    \label{w_up}
\end{equation}
where $\eta$ is known as the \textbf{learning rate} and determines how large of a step we take. In practice this process is best repeated for a small piece of the data set, which we call a \textbf{batch}, until all data have gone trough the architecture. This is a specific way to update weights called batch gradient descent.

Before we move on to some practical aspects of training in the next section, notice for a moment how conceptually similar the notion of minimizing the loss function is to the backbone of physics, the \textit{action principle}. Indeed, this is precisely what one does; we are looking for the weight matrix which minimizes the loss function. Furthermore, with our analogy to the principle of least action, it doesn't hurt to think of the weights as some coordinates in our higher dimensional space. Thus we can imagine that the process of training is tracing out a geodesic on the loss landscape.

\subsubsection*{Nuances of Training}

In many ways, training machine learning algorithms can become as much art as it is a science. Even with the best state-of-the-art architecture, a poor training regiment may fail to realize the full potential of a neural network. As such, we feel it imperative to delve into some practical aspects of training the one should have in mind. We stress that there are many facets to training and that we will focus on the handful that we feel are most critical.\footnote{A great resource for more information on training is ~\cite{goodfellow2016deep}.}

It has certainly become clear by now that there are various pieces of a neural network that we have the freedom to adjust, these are known as \textbf{hyperparameters}. Some examples of hyperparameters include the number of layers in the network, number of nodes in a layer, learning rate, batch size, and number of epochs. The optimal choice of hyperparameter is often not well understood theoretically, indeed machine learning theory is very much in its infancy, thus one must resort to trial and error or well known empirical results. 

One hyperparameter that we have already touched on briefly is the depth our network. Generally speaking, the addition of more depth to ones architecture can increase performance - this certainly seems sensible. For example, a fully connected neural network with five layers is expected to have superior performance over a similar two layer architecture. That said, it is possible to have too many layers. An architecture with a complexity that exceeds that of the data set and problem with which its tasked can succumb to \textbf{overfitting}. Put simply, the architecture in this situation has effectively learned the problem that its was presented with perfectly. At first this may seem like a positive result but, in fact, this is a terrible outcome as this architecture has simply memorized data it was trained with and lacks the ability to generalize to new data. Imagine a neural network trained to classify cats and dogs achieved 100\% accuracy on its training set due to overfitting; if it is then presented with a new data set, also of cats and dogs, it will get the correct classification label 50\% of the time. This is clearly not desirable.

Another critically important hyperparameter that one must keep in mind is the learning rate. Recall that this controls how large of a change we make when updating our weight matrices - this is $\eta$ in Eq. \ref{w_up}. Our goal is to find the minimum of the loss function but it turns out that in general there can be very many local minima in the loss landscape where one can become trapped. For this reason it is important that one does not start out with a learning rate that is too small. Indeed, imagine the scenario where one begins training in a local minima of the loss landscape. If our learning rate is too small we will never be able to take a large enough step to overcome the loss ``barrier'' that surrounds us. If we were then to increase the learning rate in this same situation, there would be no difficulty in taking a step large enough such that we could escape the local minima. Naturally, we also have the reverse scenario. If our learning rate is too large we won't have to worry about being stuck in a local minima but we may find it difficult to converge to the global minimum.  Given the two extremes it is generally good practice to dynamically adjust ones learning rate; starting with a larger value one can avoid local minima at the start of training which can then be complemented with a smaller learning rate to hopefully work ones way, in a controlled fashion, to the global minimum of the loss function.

An easy hyperparameter to overlook in training is the batch size. Recall from earlier that batch size is closely related to the amount of times the weights of the architecture are updated. The impact of small batch size is that for a given epoch we update the weights more frequently, so changes are more susceptible to outliers in the data set which may make training more difficult, i.e. updates are noisier and training can take significantly longer. The benefit is that you avoid getting stuck in local minima due to the noisiness of the updates. The extreme case is updating weights after each data point, this is known as {\bf stochastic gradient descent}. In contrast, larger batch sizes have the benefit of averaging features of a data set but this, of course, has the downside of missing subtle and legitimate features. This also has the undesirable trait that one is more susceptible to getting stuck in local minima relative to stochastic gradient descent. The extreme case here is updating the weights where the batch size is the entire data set, this is ``proper'' {\bf gradient descent}. While either of these approaches may be an appropriate training program in some cases, an intermediate choice of batch size is generally a better option; this approach, where weights are updated after a batch, is {\bf batch gradient descent}. 

Before moving on, notice that there can be some back and forth between the learning rate and batch size; a very small batch size can help one to escape local minima but this could be offset by a small learning rate. Thus, one should be careful when changing two or more hyperparameters that they understand the cumulative impact of ones changes. 

The last hyper-parameter that we will touch on is the number of epochs. When a neural network is being trained on a set of training data, the full set of data is often run through the network multiple times. Each run through of the data is called an \textbf{epoch}. If there are too few epochs, the network may not be trained well enough. If there are too many epochs, however, the network is at risk of overfitting. One can make sure that overfitting is not occurring by tracking the loss for data \textit{not} used to update weights, this is called \textbf{validation data}. If the training loss continues to decline but the validation loss has stagnated it is likely that ones model has started overfitting.

We conclude our discussion of various aspects of training neural networks by discussing \textbf{data augmentation}. This method constitutes making several copies of ones training data in such a way that the fundamental information of the data is unchanged. For example, if ones data set consists of images of cats we can augment our data set with additional copies of the \textit{same} images but where we have rotated, zoomed, flipped, added noise, etc. There are two main reasons one would want to implement such a procedure. The first is in the case that one has a limited supply of data and requires extra realizations of the same class to help an algorithm to converge. Indeed, two data of the same cat that are translated or reflected by some amount relative to each other still amounts to a cat. The final reason that data augmentation can be helpful is that it is useful for the architecture to see the same image from various angles. For example, we would like a neural network to still correctly classify an image as containing a cat even it were upside down. Indeed, if you train a fully connected neural network on images of cats that are right side up it will fail to correctly classify images of upside down cats. This is because basic neural networks are not invariant to reflection in an image. If we had instead used reflections as a form of data augmentation in our training we could encode this invariance in the weights directly. We will see in the next section an architecture that has \textit{translational} invariance directly built into its structure.

{\it Accompanying this article is a \href{https://colab.research.google.com/drive/1pt4ijRlZMtg4rUO59-gn_B82-HtyKFoH?usp=sharing}{\textcolor{magenta}{notebook}} which explicitly trains a simple neural network with gradient descent.\footnote{We also include the code in the Appendix.} We encourage the interested reader to experiment with the code.}

\section{Convolutional Neural Networks}\label{CNN}
A \textbf{Convolutional neural network} (CNN) is a modified version of a standard neural network that contains a set of layers called convolutional layers ~\cite{conv_NN,goodfellow2016deep} that serve as a {\bf feature extractor}. The main appeal of a feature extractor is that one can take a very high dimensional data set, say an image with $N$ pixels, and arrive at a compressed representation under a map $\mathbb{R}^N \rightarrow \mathbb{R}^d$ where $d \ll N$. The output can then be passed on to a fully connected neural network where inputs can now be thought of as high level features; imagine nodes representing colors, patterns, ear, nose, etc. for our cat and dog example. This can be especially helpful if your feature extractor is able to leverage redundant symmetries in your data set. This is precisely what convolutional layers are able to do as they are inherently invariant to translations.\footnote{Other implementations, such as equivariant neural networks and scattering transforms, are able to leverage even more complicated symmetries, from rotation to diffeomorphisms \cite{2019arXiv190204615C,2012arXiv1203.1513B}.}

\begin{figure}[h]
    \centering
    \includegraphics[width=0.4\columnwidth]{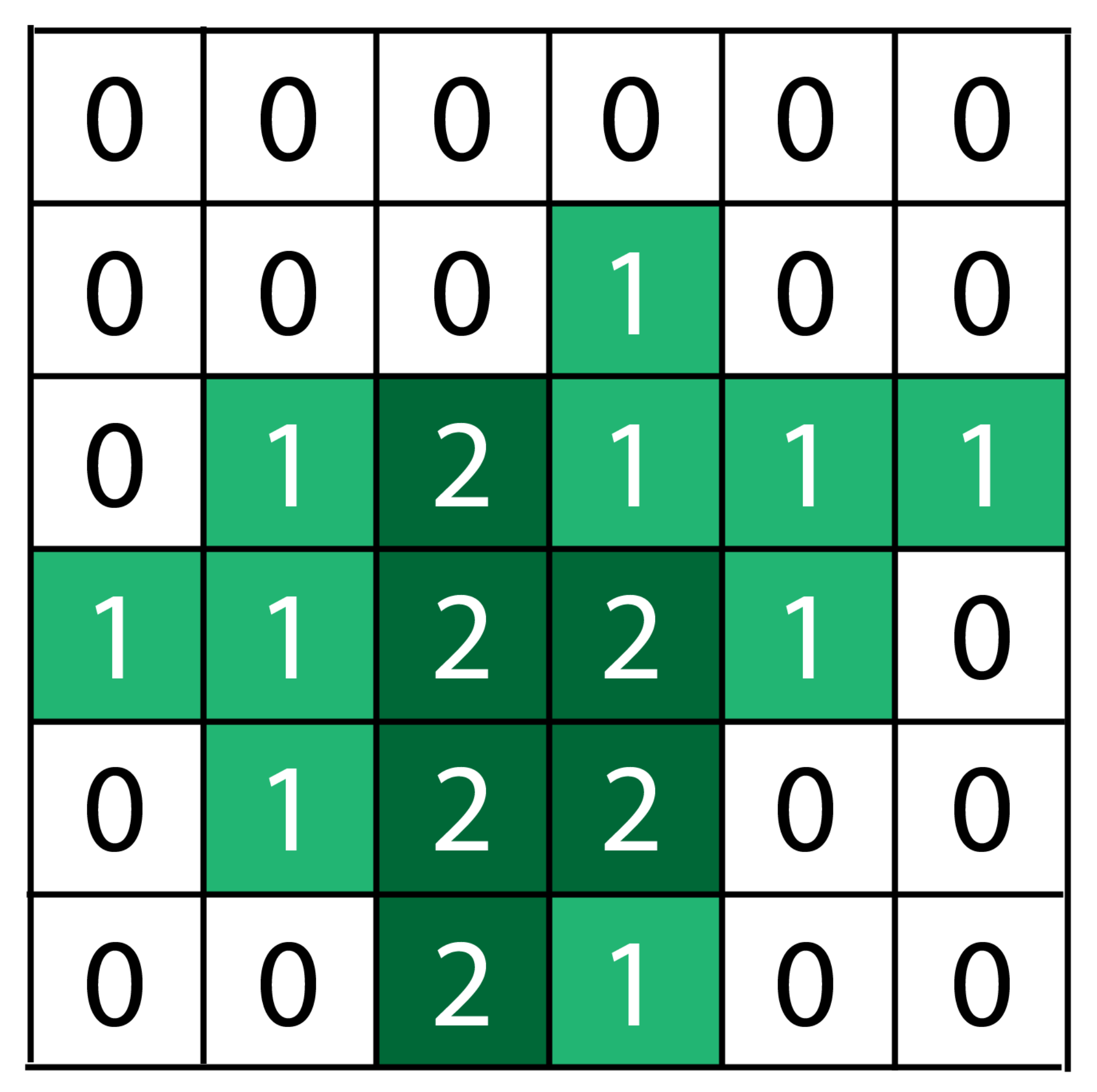}
    \caption{Our example image with pixel intensity labeled numerically and by color.}
    \label{fig:pixels_labeling}
\end{figure}

\begin{figure}[h]
    \centering
    \includegraphics[width=0.65\columnwidth]{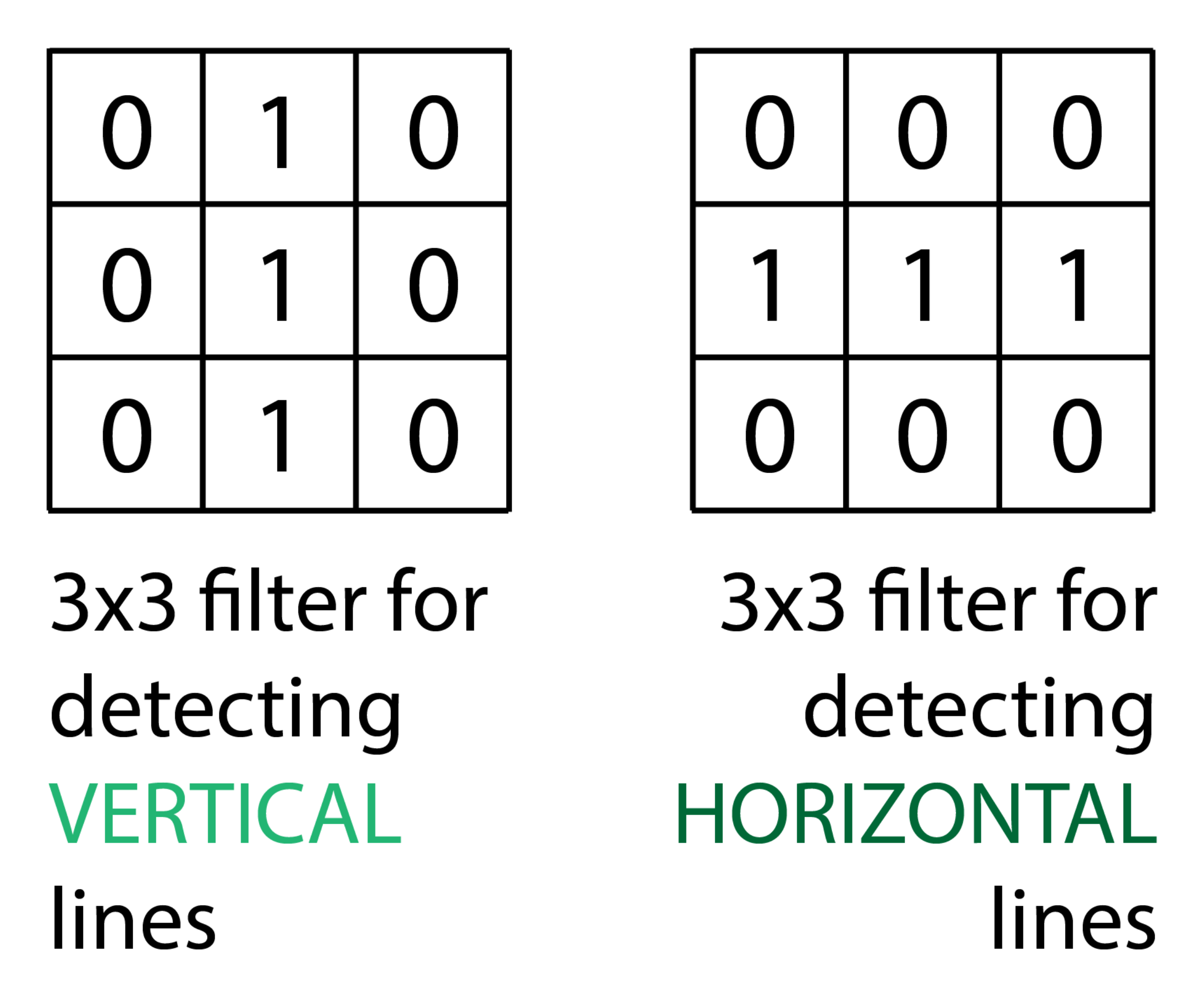}
    \caption{Two examples of $3\times3$ filters which can be used to highlight vertical and horizontal features in images.}
    \label{fig:filter_examples}
\end{figure}

\begin{figure}[h]
    \centering
    \includegraphics[width=0.8\columnwidth]{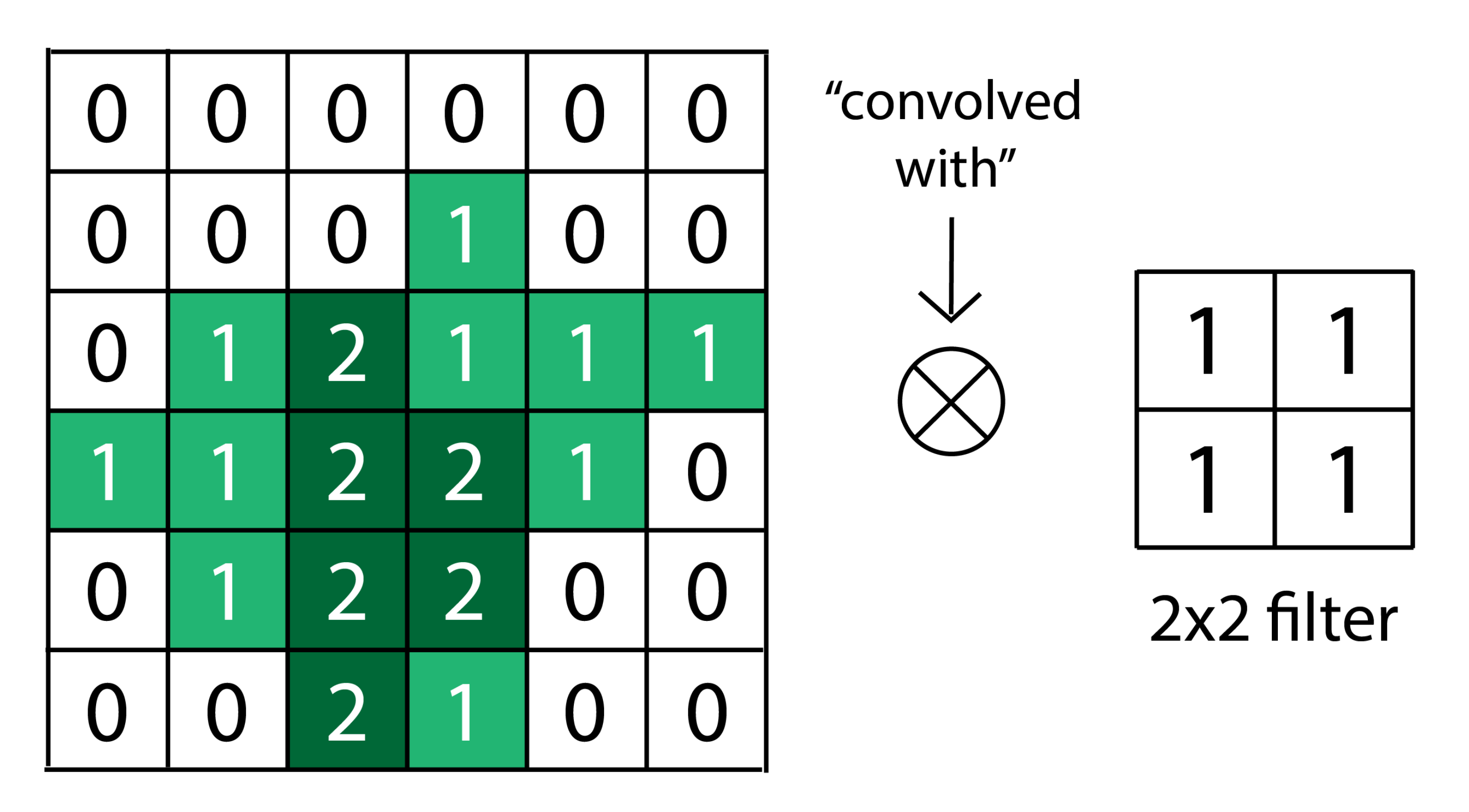}
    \caption{A $6\times6$ matrix being convolved with a $2\times2$ filter.}
    \label{fig:filter}
\end{figure}

\begin{figure}[h]
    \centering
    \includegraphics[width=0.9\columnwidth]{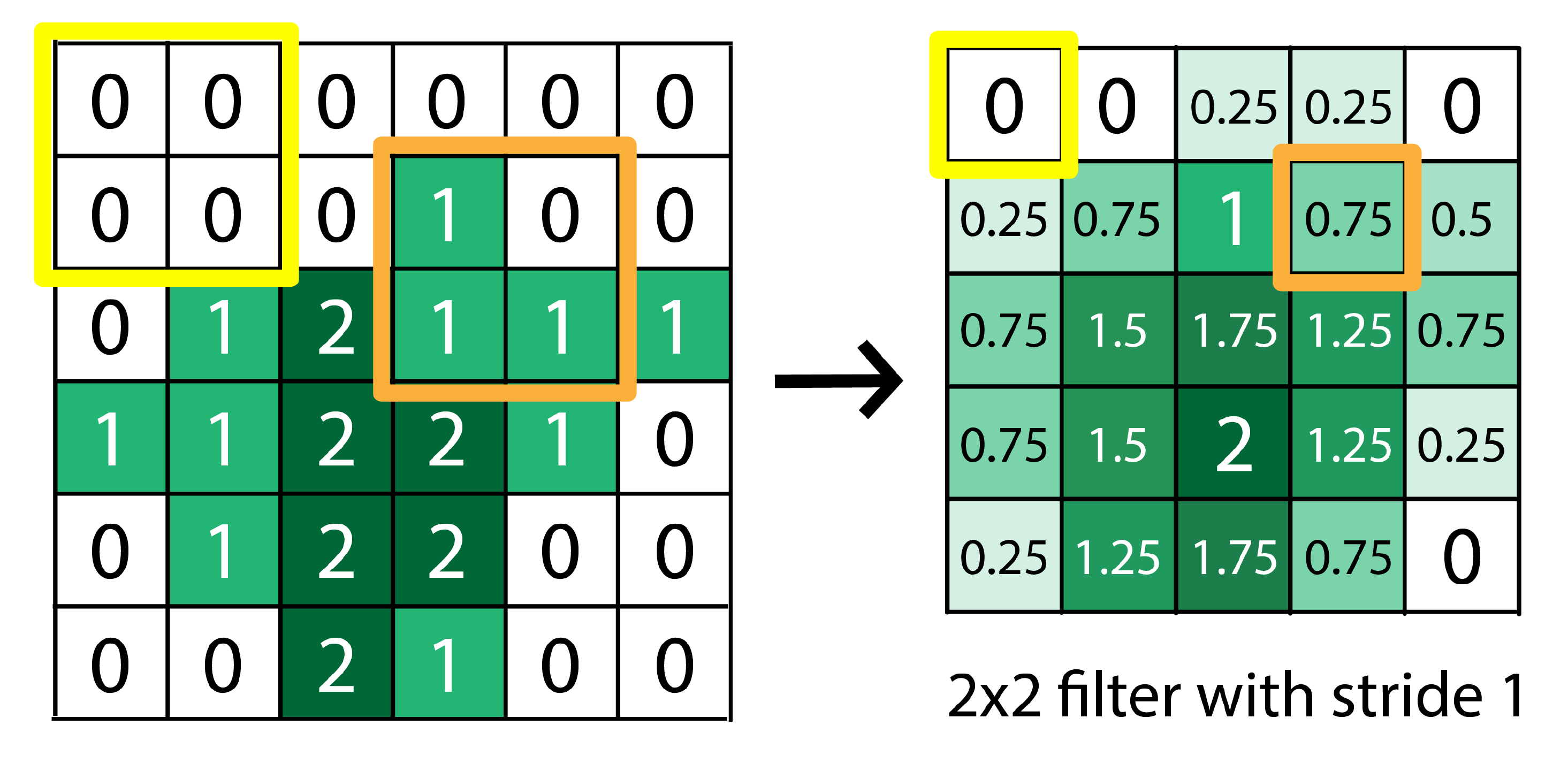}
    \caption{Convolving the filter with our image moving 1 row and 1 column at a time we end up with a $5\times5$ matrix.}
    \label{fig:stride1}
\end{figure}

\begin{figure}[h]
    \centering
    \includegraphics[width=0.8\columnwidth]{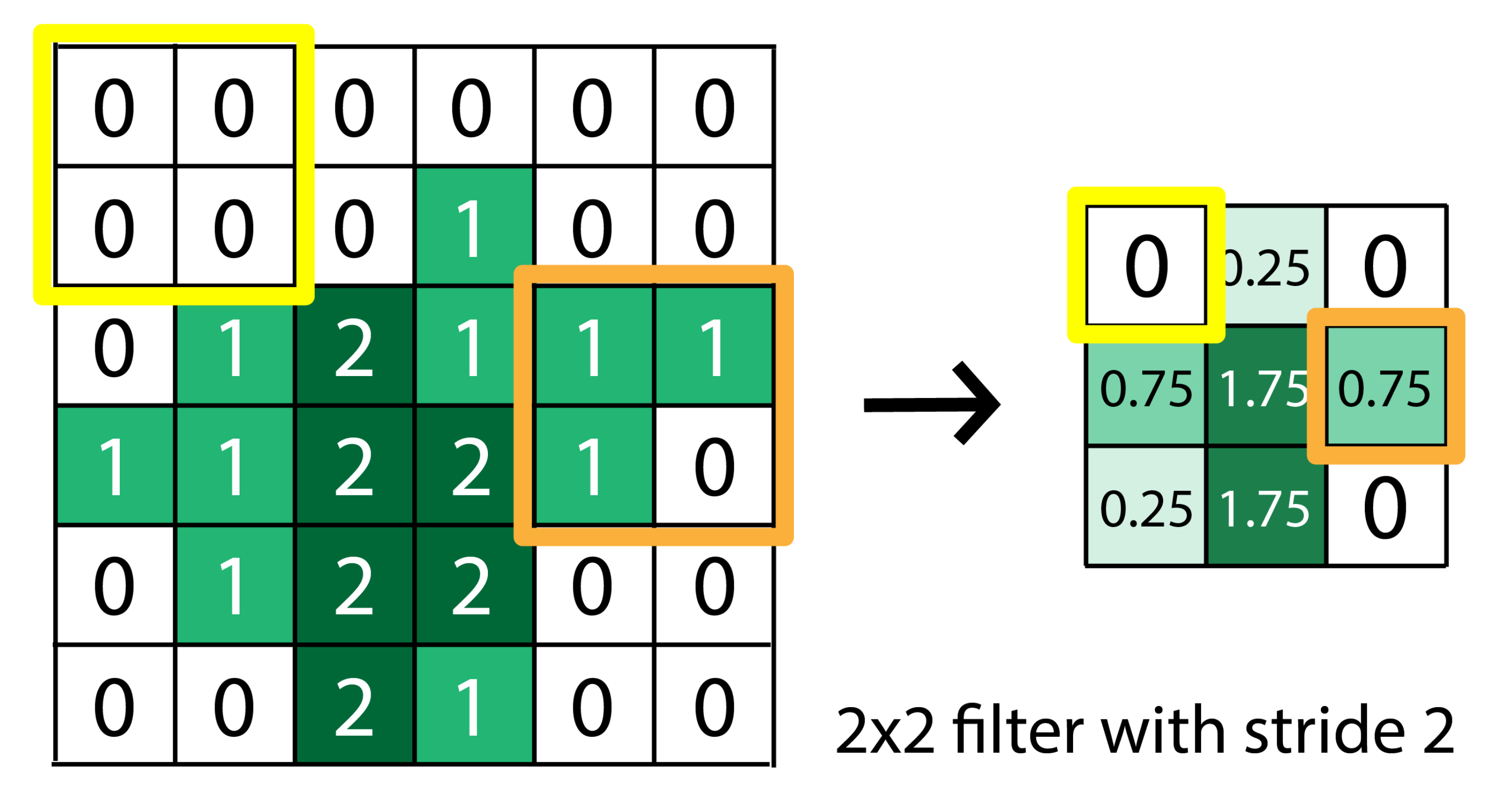}
    \caption{Same as Fig. \ref{fig:stride1} but now with a stride of 2. Notice this means the dimension of the output is smaller since it takes less moves to cover the image with the filter.}
    \label{fig:stride2}
\end{figure}

We think it best to demonstrate exactly how convolutional layers work with an explicit example. Imagine that we have the following $6\times6$ image (Fig. \ref{fig:pixels_labeling}) where the two shades of green represent pixel intensity.
We now take the convolution of our image with a {\bf filter}. The filter is the backbone of the convolutional layer; it is a small matrix and its convolution with an image serves to emphasize features consistent with the morphology of the filter, e.g. a horizontal filter will emphasise horizontal lines and wash out vertical lines. In Fig. \ref{fig:filter_examples} we show examples of what these might look like for filters of size $3\times3$.
Let us return to our example. In Fig. \ref{fig:filter} we now have our have our image and a $2\times2$ filter represented by a matrix.
As we convolve the filter with our image we then translate the filter across the image, the size of this step is the {\bf stride}.  In Fig. \ref{fig:stride1} we show the operation of our $2\times2$ filter convolved over our image with a stride of 1. Starting at the upper left corner, across the image, and then down, we end up with a $5\times5$ image. The yellow and orange boxes show the explicit mapping for two different convolution operations.
Notice that we have decreased the size of the image. In fact, had we chosen a stride of 2 we could have made the output even smaller as we show in Fig. \ref{fig:stride2}.
It is also possible to make it so that after convolution our output is of the same size. We can do this by adding null valued pixels around our image; this is called {\bf padding}.

After the convolutions the output is typically put through an activation function, e.g. leaky ReLU, and then passed on to the final part of the convolution layer, {\bf pooling}. The main goal of pooling is data compression and the two common approaches are {\bf max pooling} and {\bf average pooling}. In the former, similar to the operation of convolution, one goes over an image, say $2\times2$ pixels at a time, and maps the \textit{maximum} pixel value to the output. For the latter, one does the same but takes the average value of the four pixels to map to the output. Notice how this works as a feature extractor. If a feature consistent with the filter is present in the data, one expects signal to survive after convolution. Pooling then rewards the presence of signal, helping it to propagate. One can then compose multiple convolutional layers together such that even higher level features can be abstracted.

\begin{figure}
    \centering
    \includegraphics[width=\columnwidth]{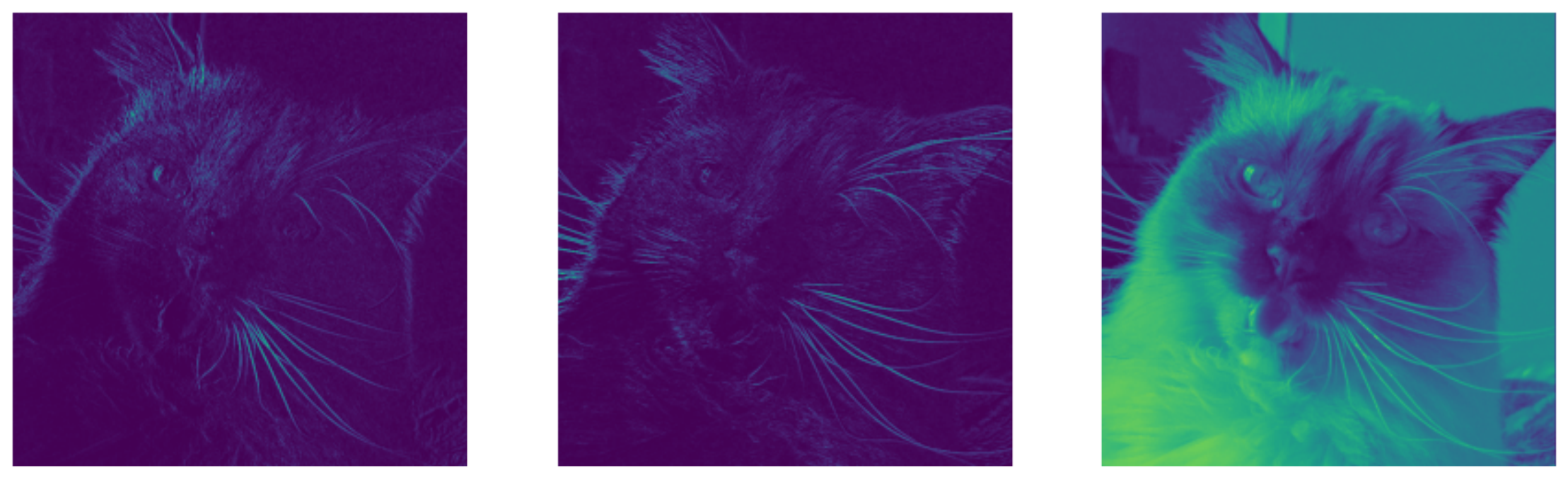}
    \includegraphics[width=\columnwidth]{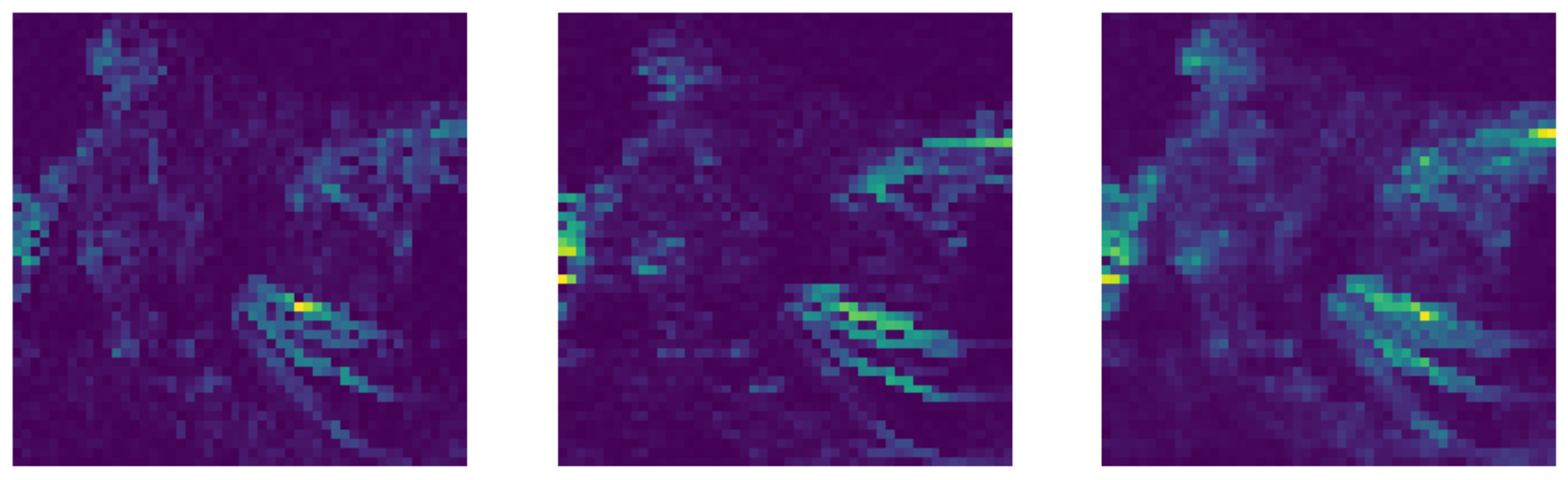}
    \includegraphics[width=\columnwidth]{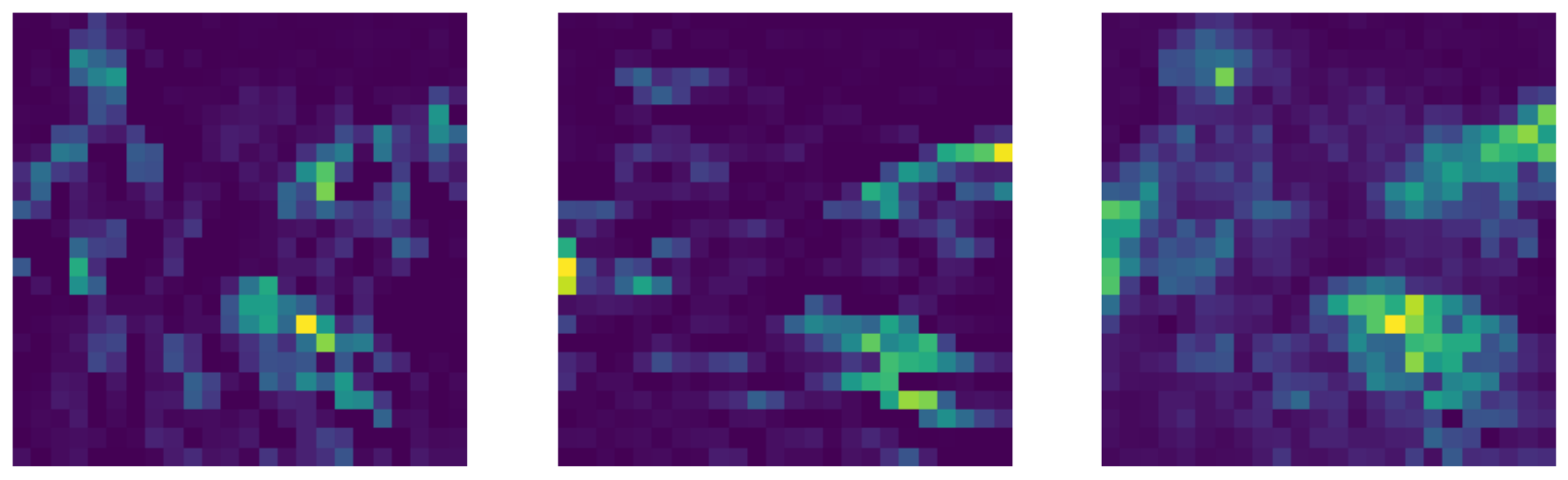}
    \caption{Features at various stages of passing through an architecture with multiple convolutional layers. We display the result when applied to the same test image for convolution with, from \textit{left to right}, vertical, horizontal, and $X$-shaped filters. After the first layer, filters convolve the output of the previous layer's convolution with the \textit{horizontal} filter.}
    \label{fig:conv_feature_maps}
\end{figure}

For some intuition into how convolution and pooling work together to extract features, we show the explicit example of the  convolution of vertical, horizontal, and $X$-shaped filters, followed by ReLU activation, and  max pooling in Fig. \ref{fig:conv_feature_maps} at the output of successive convolutional layers (notice the resolution decreases down the rows). After the first layer, filters convolve the output of the previous layer's convolution with the \textit{horizontal} filter. What should be striking in this example is how well the filters are at picking up various features of the cat. For example, we can tell the middle column is the horizontal filter as it does the best job at propagating horizontal features, e.g. top of the right ear and the whiskers. The vertical filter also has some success but mostly on defining the outline of the cat and, to a lesser extent, also the whiskers. The last filter on the other hand, a sort of superposition of the other two, struggles at first, likely due to the higher resolution, but eventually finds some relevant features inherited from the horizontal filters from earlier layers.

These two layers---convolutional and pooling---are often repeated multiple times before their final output is flattened (recall the output is still a matrix) and fed into a normal, fully connected network. Note that the filters are like weights of a fully connected neural network; they are something that must be trained. So our example above with horizontal, vertical filters was a qualitative example. During training, the algorithm will converge to a filter morphology that best suits its data set and training task.

\section{Autoencoders}\label{AE}

\begin{figure}[t]
    \centering
    \includegraphics[width=\columnwidth]{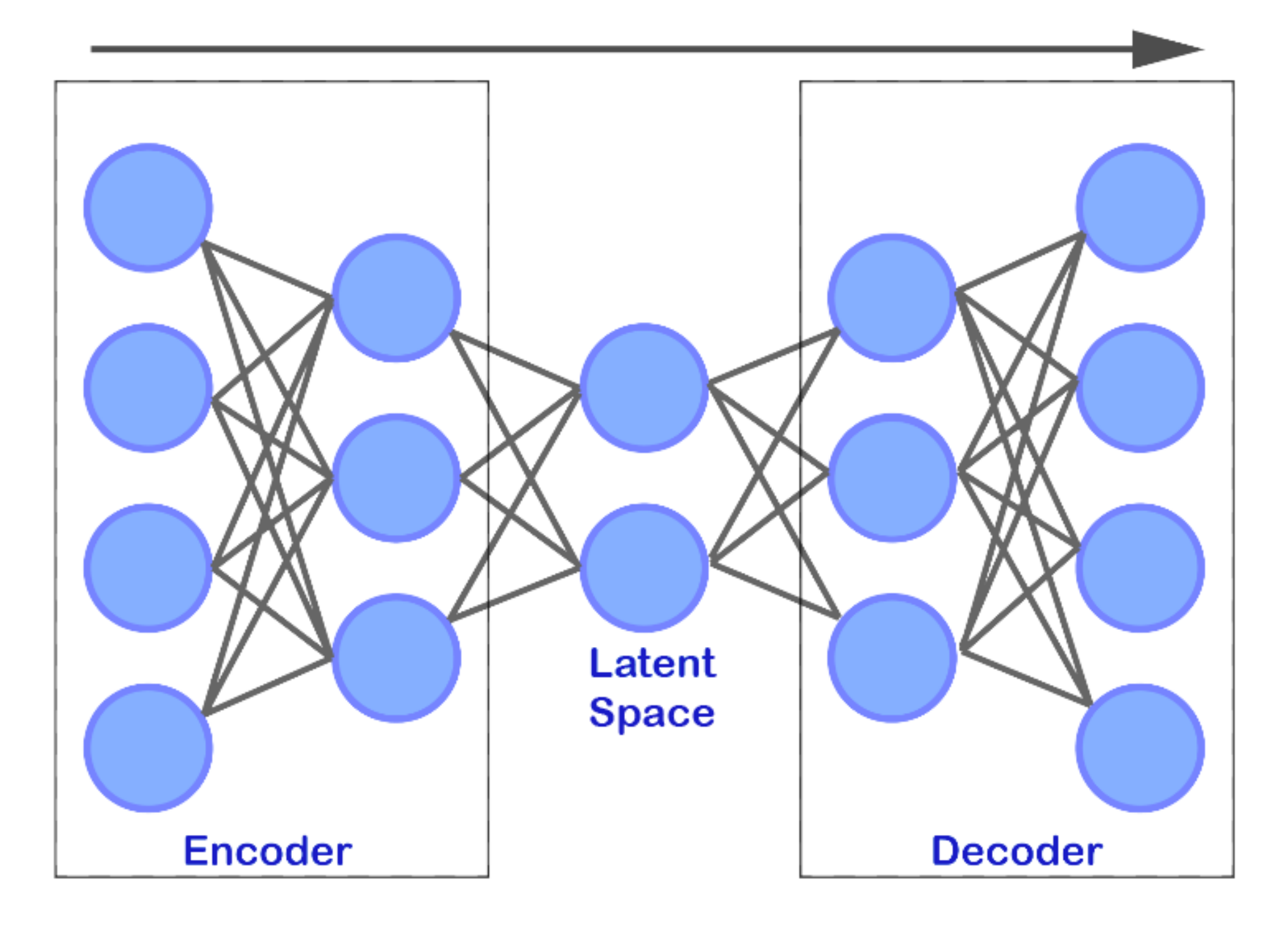}
    \caption{The basic structure of an autoencoder.}
    \label{fig:autoencoder}
\end{figure}

The last neural network based architecture we wish to introduce in this paper is the autoencoder. The \textbf{autoencoder} is a neural network that forces a data set through a bottle neck in such a way that a minimal amount of information is lost. The fundamental structure of this architecture is very similar to a standard fully connected neural network. In Fig. \ref{fig:autoencoder} we can see its three main components - the encoder, decoder, and latent space. For a simple autoencoder, the \textbf{encoder} is no different from a regular neural network, i.e. an input is taken and its dimensionality is gradually reduced from layer to layer. At the interface of the encoder and the decoder is the latent space. The \textbf{latent space} is our bottleneck, it represents a compressed representation of the data. The final part, the \textbf{decoder}, is structurally similar to the encoder but it instead \textit{increases} the dimensionality of its inputs, in this case the latent vector. A good analogy to the inner workings of an autoencoder is a zip-file. We can thing of the encoder as the algorithm which produces the zip-file, which in our analogy is the latent vector, which can then be \textit{extracted} by the decoder. Naturally we would want to limit the amount of information that is lost in this compression. For this reason, the autoencoder is a type of unsupervised algorithm - the goal is to match the input to the output as closely as possible. 

Practical applications of autoencoders include data noise reduction, data generation, and anomaly detection. Noise reduction is pretty easy to wrap ones head around. The bottle neck effectively forces the architecture to integrate out irrelevant information, so the reconstruction produced by the decoder does not realize noise at the output. This makes sense because the information contained in the structure of the data (noise is inherently \textit{not} structured) is what is most \textit{relevant} in reconstructing the input. Our discussion of neural networks to this point has been limited to deterministic architectures but it is also possible, similar to RBMs, to train an autoencoder to learn a probability distribution. One such example is a \textbf{variational autoencoder} which learns a probability distribution over the latent space. This is useful in that once the architecture is trained we can sample from that distribution, passing random values up through the decoder and produce realizations from the class the architecture was trained on.\footnote{An architecture that is similar to this is the generative adversarial network (GAN).} A final practical application of autoencoders is for \textbf{anomaly detection}. The basic premise of anomaly detection is that an architecture can be trained on a set of data in such a way that it is able to correctly reconstruct \textit{normal} data but any outliers, i.e. anomalous data, will result in poor reconstruction. This poor reconstruction, which can be measured with a loss function comparing output to input,\footnote{This is called the reconstruction loss.} is a result of the bottleneck; the architecture has effectively been trained to be an expert at reconstructing \textit{normal} data, so any data which has new or different features will fail to properly be represented in the latent space and, by extension, result in poor reconstruction at the output. One can then implement an anomaly detector by looking at the distribution of reconstruction losses for the data set and setting a threshold, in terms of loss, above which data are flagged as anomalous. A practical application could be something like a spam detector for emails.

Before concluding our discussion of autoencoders, we would like to discuss how one can interpret the latent space. Let us first take a minute to reflect on the similarities between the use of a bottleneck in an autoencoder and  physics. That said, recall our previous discussion on the connection between neural networks and the renormalization group flow. In that scenario we have the picture of reducing the number of degrees of freedom from layer to layer, naturally keeping those which are relevant. In this way, one can think of the latent space as encoding a notion of the relevant degrees of freedom for the system. If, for example, we were to train an autoencoder on the dynamics of a simple harmonic oscillator, a node in the latent space may encode information about the spring constant for the system.

\begin{figure}
    \centering
    \includegraphics[width=0.95\linewidth]{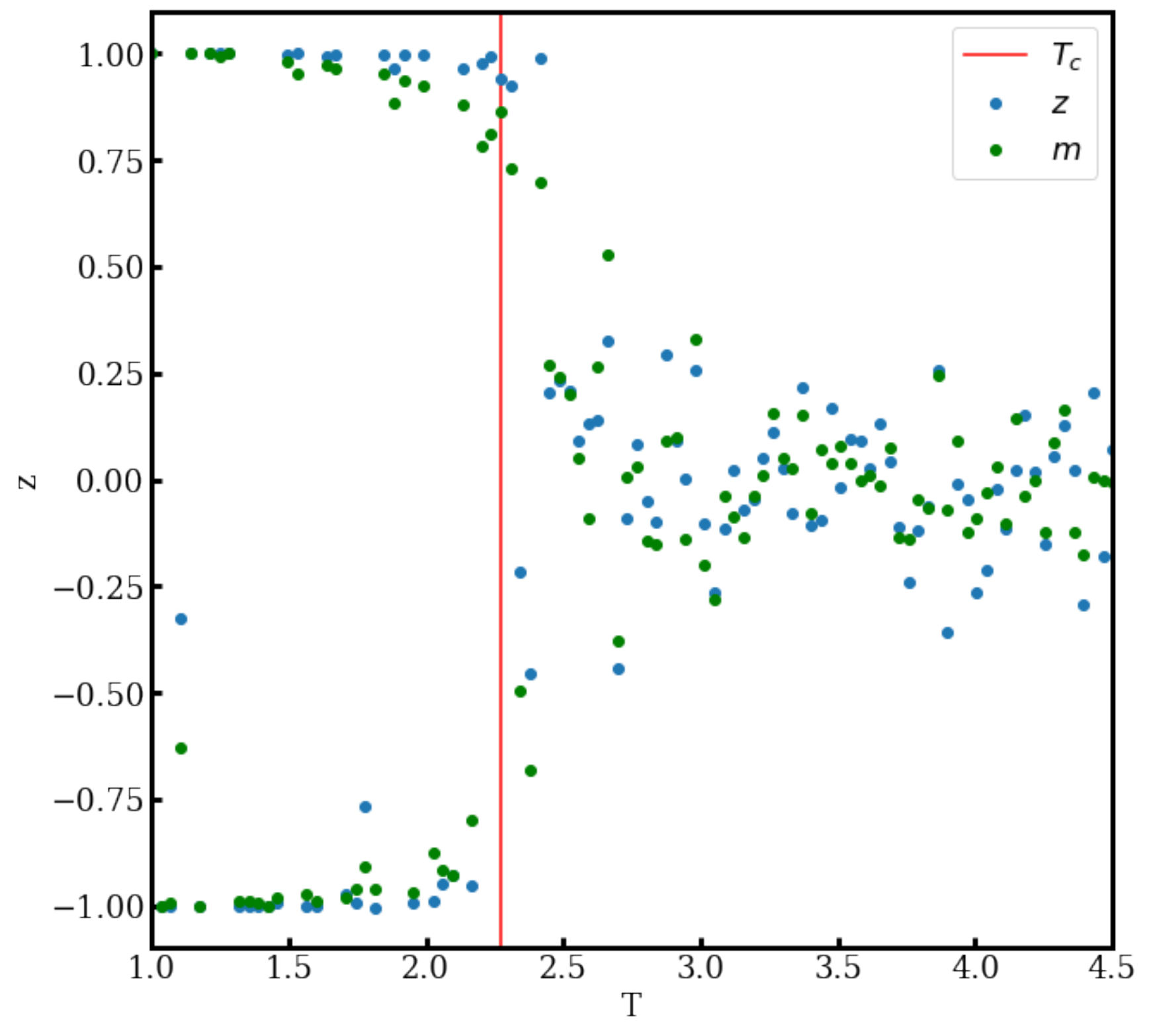}
    \caption{Plot of known order parameter (green) as function of temperature for a 2D Ising model and value at latent space (blue) for same simulations passed through an autoencoder. Red line represents the location of the critical temperature.}
    \label{fig:ising_auto}
\end{figure}

Coming full circle, let us show an explicit example where this type of architecture can be used to understand the physics of phase transitions of the Ising model. As we reviewed earlier, as the temperature in the Ising model decreases below a critical temperature, $T_c$, the system undergoes a transition from a disordered state to an ordered one. This is best understood in terms of the magnetization, which is the order parameter for the system, which we could obtain from solving Eq. \ref{mag}. A particularly insightful application of autoencoders, that can give us an intuition for what occurs in the latent space more generally, is to train one to reconstruct spin configuration of the Ising model across the phase transition.\footnote{This idea was first explored in \cite{Alexandrou_2020}.} Amazingly, training a simple fully connected autoencoder, with a latent space containing a \textit{single} neuron, for the 2D Ising model one finds that the architecture is able to discover the phase transition. We show this result in Fig. \ref{fig:ising_auto} where we plot the known magnetization (green data) from our simulations against the value of the corresponding simulation at the latent space (blue data) of our trained architecture. Notice what this means; the autoencoder has \textit{found} the relevant degree of freedom for the system without our intervention. One interpretation of this result is that the architecture effectively learned addition, i.e. to add up the spins of the lattice, which is exactly what we would expect for the magnetization.

In this example it was relatively simple to identify the single node latent space with a known quantity. In general, this will not be as straightforward. For example, with the simple architectures presented here it is possible that one can have non-trivial correlations between nodes. For example, when data for a harmonic oscillator is used to train an autoencoder with a latent space of dimension 2, one will not generally be able to identify a node with the mass or spring constant. It is possible to get around this where one adds a new term to the loss to minimize the correlation between nodes of the latent space. This new term in the loss can be thought of a Lagrange multiplier in that it adds a constraint on the system.\footnote{The addition of terms like this to a loss function is quite common and is generally referred to as regularization.} This type of architecture is called a {\bf $\beta$-variational autoencoder}. Nonetheless, we stress that it may not always be possible to find a map between the latent space representation and a known quantity but we hope our example gives one an intuition for the machinery of an autoencoder.

\section{Conclusion}\label{Con}

To appreciate the deep and profound connection between machine learning and physics let us now briefly review what we have covered. We began with a discussion of Hopfield networks and (restricted) Boltzmann machines and their deep connection to statistical mechanics. In fact, we showed that the dynamics for BMs is, for all practical purposes, indistinguishable from the Ising model. More generally, we showed that establishing a well defined quantity analogous to a Hamiltonian, where we critically restricted the weights to be symmetric, led to desirable dynamics. Concretely, we saw that the repeated application of the dynamical equations for the Hopfield network, Eq. \ref{eq:Hopp-dyn}, helped it to converge on stored \textit{memories} which can be thought of as attractors for the system. This is all well and good, but in practice Hopfield networks and Boltzmann machines are either not particularly useful or very difficult to train relative to other architectures. Their simple structure, however, is useful to understand general feature of machine learning. 

Using our \textit{intuition}, developed from energy based machine learning algorithms, we then delved into a few architectures related to the backbone of modern machine learning - the feedforward neural network. Considering the flow of information from layer to layer we stumbled upon the idea that neural networks implement coarse graining between layers, integrating out irrelevant information and propagating that which is relevant, which is a realization of a renormalization group flow within a machine learning architecture. We then saw how one can draw a direct analogy between minimizing the loss function for a neural network and the action principle, this then led us to imagine training tracing out geodesics along the loss landscape. In the context of convolutional neural networks we saw that leveraging symmetries (you can also think of this as redundancies) in the structure of the architecture can greatly improve the performance. This naturally resonates with the notion of gauge and diffeomorphism invariance which are the backbones of field theory and general relativity. Lastly, we saw that an autoencoder, which realizes a notion of an RG flow, was able to explicitly detect the phase transition in the Ising model without \textit{any} outside input.

We hope that those initially skeptical are convinced of this deep and fundamental connection. As we mentioned at the beginning of this article, our hope is not only that we may garner a deeper understand of machine learning through the lens of theoretical physics (indeed there is much work to be done in developing the theory of machine learning) but that machine learning may also inspire new discoveries in physics. In this sense we don't necessarily mean, what is now quite common in the field, the application of machine learning algorithms to a physical data set, but rather new ways of thinking about physics. Concretely, if we are to take these connection seriously, where and how does a notion of learning, training, etc. manifest itself in the laws of physics?

 \section*{Acknowledgements}
The authors thank Sergei Gleyzer, Zach Hemler, and Stefan Stanojevic for insightful comments on an early draft of this work.

\bibliographystyle{unsrt}
\bibliography{bibo.bib}

\onecolumngrid
\appendix
\newpage
\section*{Explicitly Coded Neural Network Example}

The following code executes the same math done in \textit{The Physics of Machine Learning: An Intuitive Introduction for the Physical Scientist}. It utilizes the well known MNIST (Modified National Institute of Standards and Technology) dataset which is a set of 70,000 images of handwritten digits 0 through 9. 60,000 of the images are designated for training and 10,000 are designated for testing. This example will utilize the training and testing MNIST data in order to teach our basic neural network architecture to read handwritten numbers.

\vspace{3mm}

\noindent Importing packages:
\begin{python}
import torch
import numpy as np
import matplotlib.pyplot as plt
from keras.datasets import mnist
\end{python}

\vspace{3mm}

\noindent The following lines of code handle downloading the MNIST dataset, and opening the corresponding file. This is how we obtain variables $x\_train$, $y\_train$, $x\_valid$, and $y\_valid$. Note that the x variables represent the input, and the y variables represent a key to check the accuracy of the output.

\begin{python}
(x_train, y_train), (x_valid, y_valid) = mnist.load_data()

dtype = torch.float # set the data type to float, to preserve decimal places
device = torch.device("cpu") # set device to cpu to optimize runtime

# input_size and output_size are characterstic to the mnist dataset, so do not
# alter them, or else this network will not run correctly (if it all)
input_size = 784 # input size is 28 x 28 = 784, which is charactersitic to the mnist dataset
output_size = 10 # output size is 10 since there are 10 possible digits that the input could be (0-9)
\end{python}

\vspace{3mm}

\noindent Hyperparameters (feel free to alter in the link to the google colab and see how the result is affected):

\begin{python}
batch_size = 64
hidden_size = 100
num_epochs = 2
learning_rate = 1e-3
\end{python}

\vspace{3mm}

\noindent Making the MNIST data the right format:

\begin{python}
x = torch.from_numpy(x_train)

x_valid_torch = torch.from_numpy(x_valid)

y = torch.from_numpy(y_train)

y_valid_torch = torch.from_numpy(y_valid)
\end{python}

\vspace{3mm}

\noindent Turning the output validation data into ``one-hot encodings.'' This encoding is where we take the digits 0 through 9 and equate them with unit vectors in nine dimensional space, so 0 $\rightarrow$ (1,0,...,0), 1 $\rightarrow$ (0,1,...,0), ... , 9 $\rightarrow$ (0,0,...,1). This designation allows the algorithm to check the output more easily.

\begin{python}
y_train_hot_torch = torch.nn.functional.one_hot(y.to(torch.int64))

y_valid_hot_torch = torch.nn.functional.one_hot(y_valid_torch.to(torch.int64))
\end{python}

\vspace{3mm}

\noindent Next we initialize random weight matrices. Note that the sizes must be compatible with our network.

\begin{python}
w1 = torch.randn(input_size, hidden_size, device=device, dtype=dtype)
w2 = torch.randn(hidden_size, output_size, device=device, dtype=dtype)
\end{python}

\vspace{3mm}

\noindent Lastly we have the main architecture. In the following code, we will complete a forward pass. (The ongoings of said forward pass will be mostly explained here, prior to the code itself, because, for the code to work properly, we should not break up the for-loops.)

\noindent First, we matrix multiply the input with first weight matrix,
$$z_1 = \mathbf{W_1}\mathbf{x}_{\rm{input}}.$$

\noindent Next, we will apply the ReLU activation function,
$$\sigma(z_1) = \text{ReLU}(z_1) = \mathbf{x_2}.$$

\noindent After, we matrix multiply our second weight matrix to this layer,
$$z_2 = \mathbf{W_2x_2}.$$

\noindent And then we do a final ReLU activation function on this layer, returning our desired output, y, (in the code this is y\_pred):
$$y_{\text{out}} = \sigma(z_2) = \text{ReLU}(z_2).$$

\noindent Without renaming variables as we did above, this is what the entire expression looks like:
$$y_{\text{out}} = \sigma(\mathbf{W_2}\sigma(\mathbf{W_1}\mathbf{x}_{\text{input}})).$$

\noindent Afterwards, we start the backpropagation process. In this process, we must calculate the loss and then apply gradient descent as described in the paper. 

\noindent Our loss function is given by
 
 $$\mathcal{L} = \frac{1}{n}(y_{\text{out}}-y_{\text{known}})^2,$$
 \noindent (note that a different loss function, known as cross entropy loss, would be more appropriate for this specific example, but for simplicity's sake we will use mean squared error (MSE))
 \noindent and of course to execute gradient descent, we must take the derivatives of the loss with respect to each weight matrix.
 \noindent Let us first differentiate with respect to the outtermost weights, $\mathbf{W_2}$, and use the chain rule to do so:

 $$\frac{\partial \mathcal{L}}{\partial \mathbf{W_2}} = \frac{\partial z_2}{\partial \mathbf{W_2}}\frac{\partial \sigma(z_2)}{\partial z_2}\frac{\partial y_\text{out}}{\partial \sigma(z_2)}\frac{\partial \mathcal{L}}{\partial y_\text{out}}.$$

 \noindent Starting with the derivatives on the right and moving to the left:

$$\frac{\partial \mathcal{L}}{\partial y_{\text{out}}} = \frac{2}{n}(y_{\text{out}}-y_{\text{known}})^\text{T}, $$

$$\frac{\partial y_\text{out}}{\partial \sigma(z_2)} = 1,$$

\begin{equation*}
\frac{\partial\sigma(z_2)}{\partial z_2} = \Bigg\{
\begin{array}{ll}
      1 ~~~ z_2 \geq 0 \\
      0 ~~~ z_2 < 0 \\
\end{array},
\end{equation*}

 $$\frac{\partial z_2}{\partial \mathbf{W_2}}=\mathbf{x_2} = \rm{ReLU}(z_1)$$

 \noindent we obtain,
 $$\frac{\partial \mathcal{L}}{\partial \mathbf{W_2}} = \Bigg\{
\begin{array}{ll}
      {\text{ReLU}(z_1)\frac{2}{n}(y_{\text{out}}-y_{\text{known}})^\text{T}} ~~~ z_2 \geq 0 \\
      0 ~~~~~~~~~~~~~~~~~~~~~~~~~~~~~~~~~~~~~~ z_2 < 0 \\
\end{array}.$$

 \noindent From the paper, we know that $\Delta \mathbf{W_2} = \frac{\partial \mathcal{L}}{\partial \mathbf{W_2}},$ and therefore we will keep this result for the weight updates.

 \noindent Next, to find $\Delta \mathbf{W_1}$, we must differentiate the loss with respect to $\mathbf{W_1}$. Again we use the chain rule:

 $$\frac{\partial\mathcal{L}}{\partial\mathbf{W_1}}=\frac{\partial z_1}{\partial \mathbf{W_1}}\frac{\partial \sigma(z_1)}{\partial z_1}\frac{\partial y_{\rm{out}}}{\partial \sigma(z_1)}\frac{\partial \mathcal{L}}{\partial y_{\rm{out}}}.$$

\noindent Then we have to take the four derivatives as required by the chain rule above:

$$\frac{\partial \mathcal{L}}{\partial y_{\text{out}}} = \frac{2}{n}(y_{\text{out}}-y_{\text{known}})^\text{T},$$
$$\frac{\partial y_{\text{out}}}{\partial \sigma(z)}= \mathbf{W_2},$$
\begin{equation*}
\frac{\partial\sigma(z_1)}{\partial z_1} = \Bigg\{
\begin{array}{ll}
      1 ~~~ z_1 \geq 0 \\
      0 ~~~ z_1 < 0 \\
\end{array},
\end{equation*}
\noindent and
$$\frac{\partial z}{\partial \mathbf{W_1}} = \mathbf{x}_{\text{input}}.$$

\noindent Putting it all together,

$$\frac{\partial\mathcal{L}}{\partial\mathbf{W_1}} = \Bigg\{
  \begin{array}{ll}
      \mathbf{x}_{\text{input}}\frac{2}{n}(y_{\text{out}}-y_{\text{known}})^\text{T}\mathbf{W_2} ~~~ z_1 \geq 0 \\
      0 ~~~~~~~~~~~~~~~~~~~~~~~~~~~~~~~~~~~~~~~~~~~ z_1 < 0 \\
\end{array}$$

\noindent And the above is our formula for $\Delta \mathbf{W_1}$.

\noindent Finally, the weights can be updated as such:
$$\mathbf{W_1}^\text{new}= \mathbf{W_1} + \eta\Delta\mathbf{W_1}$$

$$\mathbf{W_2}^\text{new}= \mathbf{W_2} + \eta\Delta\mathbf{W_2}$$

\noindent Now see all of these steps laid out in the code below:

\begin{python}
for epoch in range(num_epochs): # here we are iterating over each epoch
#(number of times we pass all of the data through the code)

  epoch_loss = 0 # we create a variable to track the loss for each epoch

  for ind in range(x_train.shape[0] // batch_size): # next we iterate over each
  # element of the training data, splitting it up into batches
    
    # create a variable to track the gradient descent of the w1 matrix
    grad_w1_batch = 0 

    # create a variable to track the gradient descent of the w2 matrix
    grad_w2_batch = 0 

    # create a variable to track the loss across the batch
    batch_loss = 0 

    for delta in range(batch_size): 
    # now we are iterating over every element in a batch

      # calculate the current batch's starting index (ind * batch_size), 
      # and then add delta to get the specific index of our current input
      curr = (ind * batch_size) + delta
      new_x = x[curr].reshape(784,1) # reshape the data from 28x28 to 784x1
      z1 = w1.mm(new_x.type(torch.FloatTensor))
      # matrix multiply the input (new_x) with our weight matrix (w1)
      # This last step corresponds to the first equation in the 
      # text block above

      # we want to use relu activation, so by utilizning clamp() 
      # and setting values below 0 -> 0, we can recreate the relu function - 
      # this corresponds to the next equation above
      relu1 = z1.clamp(min=0) # above this is written as sigma of z1

      # we'll make this definition simply for consistency of notation
      x2 = relu1

      # now we multiply the output by the second weight matrix
      z2 = w2.mm(x2)

      # next we apply the second relu activation function
      relu2 = z2.clamp(min=0)

      # and finally we redefine this value as our output
      y_out = relu2

      # compute loss and update tracker variable batch_loss
      loss = (1/10)*(y_out - y[curr]).pow(2).sum().item()

      # avoid double counting when we sum over each index in batch
      batch_loss = (batch_loss + loss) / 2 

      '''
      Here we'll start the back propagation process.

      First things first, we calcualte the derivative of the loss with respect 
      to the second weight matrix as outline by the steps above.
      '''

      # we begin by taking the derivative of the loss with respect to y_out
      # note that due to matrix calculus rules, we take the transpose
      dL_dy = (2.0/10) * (y_out - y_train_hot_torch[curr].reshape(10,1)).t()

      # next remember that the derivative of y with respect to sigma(z2) is 1 
      # so (dL/dy)(dy/dsigma(z2)) = dL/dsigma(z2) => (dL/dy) = dL/dsigma(z2)

      dL_drelu2 = dL_dy

      # To apply the relu derivative, we must take the negative z2 
      # terms to be zero:

      dL_drelu2[z2.t() < 0] = 0

      # now we'll set this adjusted dL/drelu2 equal to dL/dz2

      dL_dz2 = dL_drelu2

      # and finally we get our sought after change in w2 as we derived in the
      # math above

      dL_dw2 = relu1.mm(dL_dz2)

      '''
      Now we'll calculate the derivative of the loss with respect to the 
      first weight matrix. 
      '''

      #Since we have already derived an expression for (dL/dy), we can reuse 
      # it from above to obtain the derivative of L with respect to sigma(z1):
      dL_drelu1 = dL_dy.mm(w2) 

      # now to apply the relu derivative, we use the same steps as above: 

      dL_drelu1[z1.t() < 0] = 0 

      # doing some renaming of variables to match our mathematical derivation:

      dL_dz1 = dL_drelu1

      # and finally determining the derivative of L with respect to w1:
      dL_dw1 = (new_x.type(torch.FloatTensor)).mm(dL_dz1)

      # and now we add each of these delta w's to the total change of w's 
      # for the batch
      grad_w1_batch += dL_dw1.t()
      grad_w2_batch += dL_dw2.t()

    epoch_loss += batch_loss

    # backprop to compute gradients of w1 and w2 with respect to loss 
    # note that the weights are only updated as much as the 
    # learning_rate allows!
    w1 -= learning_rate * (grad_w1_batch)
    w2 -= learning_rate * (grad_w2_batch)
    if curr 
      print(curr + 1, epoch_loss/curr)

print("final loss:")
print(epoch_loss/curr)
\end{python}

\vspace{4mm}
You may end up with a final loss that looks like this: 0.4426840823446696. As an exercise, try implementing this code while switching to a sigmoid activation function. You may notice a much smaller loss in the end. For more practice, try altering hyper-parameters such as the batch size, number of layers, learning rate, and the number of epochs. Take note of how each adjustment worsens or improves the loss.

\end{document}